\def\draftversion{false}

\RequirePackage{ifthen}
\ifthenelse{\equal{\draftversion}{true}}{
  \documentclass[aps,prl,10pt,galley,amsmath,amssymb,showpacs,
                 superscriptaddress]{revtex4}
}{
  \documentclass[aps,prl,10pt,twocolumn,amsmath,amssymb,showpacs,
                 superscriptaddress]{revtex4-1}
}

\ifthenelse{\equal{\draftversion}{true}}{
  \marginparwidth 2.7in
  \marginparsep 0.5in
  \newcounter{comm} % counter for commentaries
  \def\commnext{\stepcounter{comm}}
  \def\commtext{{\bf\color{blue}[\arabic{comm}]}}
  \def\commmar{{\bf\color{blue}[\arabic{comm}]}}
  \def\dvm#1{\commnext\marginpar{\small DV\commmar: #1}\commtext}
  
  \def\hsm#1{\commnext\marginpar{\small HS\commmar: #1}\commtext}
  \def\mym#1{\commnext\marginpar{\small MY\commmar: #1}\commtext}
  \def\gbm#1{\commnext\marginpar{\small GB\commmar: #1}\commtext}

}{
  \def\dvm#1{}
  \def\scm#1{}
  \def\hsm#1{}
  \def\mym#1{}
  \def\gbm#1{}
  
}

\usepackage[]{cleveref}
\usepackage[]{graphicx}
\usepackage[]{verbatim}
\usepackage[dvipsnames]{xcolor}
\usepackage{tabularx}

\usepackage{amsfonts}
\usepackage{amsmath}
\usepackage{amssymb}
\usepackage{bm}
\usepackage{graphicx}
\usepackage[colorlinks=true,urlcolor=blue,citecolor=blue,linkcolor=blue,bookmarks=false,pdfstartview={FitH}]{hyperref}
\usepackage{mathrsfs}
\usepackage[lofdepth,lotdepth,caption=false]{subfig}
\usepackage{varwidth}
\usepackage{wrapfig}
\usepackage{times}
\usepackage{longtable}
\usepackage{multirow}
\usepackage{tikz}

\makeatletter
\AtBeginDocument{\@ifpackageloaded{natbib}{\ifNAT@numbers\if@filesw\immediate\write\@auxout{\string\global\string\NAT@numberstrue}\fi\fi}{}}
\makeatother

\begin{document}

\title{Covalency-driven collapse of strong spin-orbit coupling \\in face-sharing iridium octahedra}

\author{Mai Ye}
\email{mye@physics.rutgers.edu}
\affiliation{Department of Physics and Astronomy, Rutgers University, Piscataway, NJ 08854, USA}
\author{Heung-Sik Kim}
\email{hk676@physics.rutgers.edu}
\affiliation{Department of Physics and Astronomy, Rutgers University, Piscataway, NJ 08854, USA}
\author{Jae-Wook Kim}
\affiliation{Department of Physics and Astronomy, Rutgers University, Piscataway, NJ 08854, USA}
\author{Choong-Jae Won}
\affiliation{Max Planck POSTECH/Korea Research Initiative, Pohang University of Science and Technology, Pohang 37673, Korea}
\affiliation{Laboratory of Pohang Emergent Materials, Pohang Accelerator Laboratory, Pohang 37673, Korea}
\author{Kristjan Haule}
\affiliation{Department of Physics and Astronomy, Rutgers University, Piscataway, NJ 08854, USA}
\author{David Vanderbilt}
\affiliation{Department of Physics and Astronomy, Rutgers University, Piscataway, NJ 08854, USA}
\author{Sang-Wook Cheong}
\affiliation{Department of Physics and Astronomy, Rutgers University, Piscataway, NJ 08854, USA}
\affiliation{Rutgers Center for Emergent Materials, Rutgers University, Piscataway, NJ 08854, USA}
\author{G. Blumberg}
\email{girsh@physics.rutgers.edu}
\affiliation{Department of Physics and Astronomy, Rutgers University, Piscataway, NJ 08854, USA}
\affiliation{National Institute of Chemical Physics and Biophysics, 12618 Tallinn, Estonia}

\date{\today}

\begin{abstract}
We report \textit{ab-initio} density functional theory calculation and Raman scattering results to explore the electronic structure of Ba$_5$CuIr$_3$O$_{12}$ single crystals. This insulating iridate, consisting of face-sharing IrO$_6$ octahedra forming quasi-one-dimensional chains, cannot be described by the local $j_{\rm eff}$\,=\,1/2 moment picture commonly adopted for discussing electronic and magnetic properties of iridate compounds with IrO$_6$ octahedra. The shorter Ir-Ir distance in the face-sharing geometry, compared to corner- or edge-sharing structures, leads to strong covalency between neighboring Ir. Then this strong covalency results in the formation of molecular orbitals (MO) at each Ir trimers as the low-energy electronic degree of freedom. The theoretically predicted three-peak structure in the joint density of states, a distinct indication of deviation from the $j_{\rm eff}$\,=\,1/2 picture, is verified by observing the three-peak structure in the electronic excitation spectrum by Raman scattering.
\end{abstract}

\maketitle

The competition between covalency and electron correlations is a core concept in the study of the Mott physics~\cite{Imada1998}. A canonical example is the contrast between the Mott insulating and metallic behaviors of 3$d$ and 4$d$ transition-metal oxides (TMOs), respectively. In these systems, the Mott phases in 3$d$ TMOs are attributed to the smaller covalency of 3$d$ orbitals, {\it i.e.} smaller overlap integrals and the resulting stronger Coulomb repulsion~\cite{Khomskii2014}, while the enhanced covalency and weaker Coulomb repulsion in 4$d$ TMOs lead to metallicity~\cite{Cox1992,Damascelli2000,Baumberger2006}. 

An interesting twist to the above simplistic picture happens in 5$d$ TMOs, especially in iridate compounds with quasi-two-dimensional layered structures~\cite{Crawford1994,Okabe2011,Singh2010,Ohgushi2013} where insulating behavior with local magnetic moments were found. The key to this puzzle was found to be the presence of the strong spin-orbit coupling (SOC) in the Ir 5$d$ orbital. Namely, SOC introduces splitting of the broad 5$d$ bands into narrow subbands and forms spin-orbital-entangled local moments identified with the effective total angular momentum $j_{\rm eff}$\,=\,1/2~\cite{Kim2008,Kim2009}. Since then, the $j_{\rm eff}$\,=\,1/2 scenario has become a cornerstone in the study of correlated phases in 5$d$ TMOs, and various theoretical suggestions of potential novel quantum phases such as high-T$_c$ superconductivity~\cite{Kim2014,Kim2015} or quantum spin liquid phases have been made based on this picture~\cite{Jackeli2009,Witczak-Krempa2014,Rau2016}. \hsm{Comments added using the predefined command are located in the right column.} 

\begin{figure}
\includegraphics[width=0.50\textwidth]{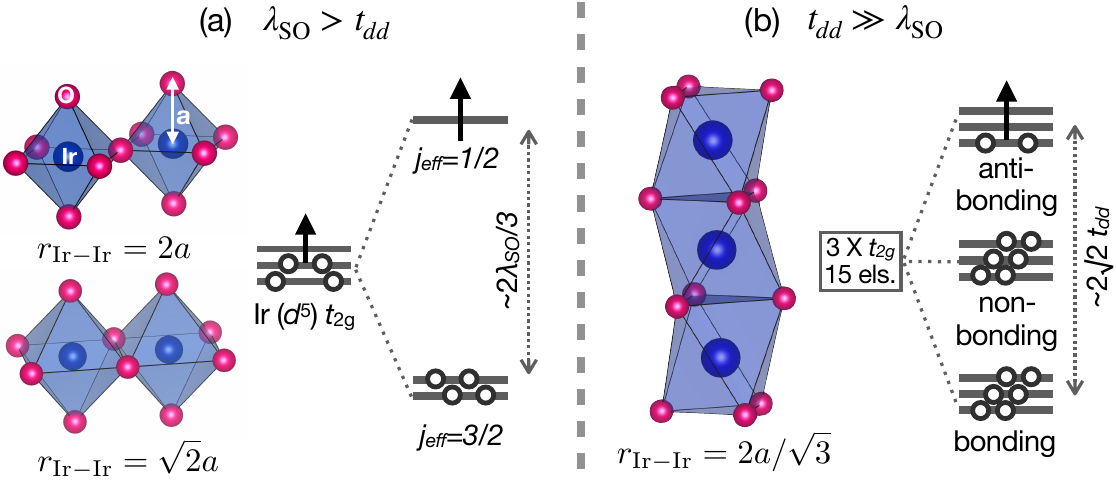}
\caption{\label{fig:Fig0}Three representative local geometries consisting of IrO$_6$ octahedra and their schematic energy-level diagrams. (a) depicts corner- and edge-sharing geometries where the size of spin-orbit coupling (SOC) $\lambda_{\rm SO}$ is larger than the covalency between neighboring Ir 5$d$-orbitals. (b) shows a face-sharing local geometry, where the Ir-O bond length is shorter compared to the other two cases so that the strength of $d$-$d$-covalency $t_{\rm dd}$ can overcome SOC. Schematic energy level diagram for each case, the conventional atomic $j_{\rm eff}$-picture and a trimer molecular-orbital (MO) picture for (a) and (b) respectively, are represented. }
\end{figure}

A critical necessary condition for the $j_{\rm eff}$\,=\,1/2 picture is the presence of (pseudo-)cubic IrO$_6$ octahedra, as shown in Fig.~\ref{fig:Fig0}(a), where the Ir $t_{\rm 2g}$ orbital ($l_{\rm eff}$ = 1) splits into the $j_{\rm eff}$\,=\,1/2 and 3/2 subspaces. Introducing non-cubic crystal fields can mix the two $j_{\rm eff}$ subspaces and break the $SU$(2) symmetry of the $j_{\rm eff}$\,=\,1/2 pseudospin. In various iridate compounds, however, such non-cubic distortions of IrO$_6$ octahedra were found to be not strong enough to qualitatively change the $j_{\rm eff}$ picture~\cite{Zhang2013,Gretarsson2013,Hozoi2014,Liu2012}, except in a small number of examples where the non-cubic distortions are exceptionally huge~\cite{SWKim2015}. Hence the belief for the validity of the $j_{\rm eff}$ scenario in general iridates has become strengthened, and it has been adopted even in situations where the applicability of the scenario is not rigorously justified~\cite{Ju2013}. 

In this Rapid Communications, we study a material in which the local $j_{\rm eff}$ moment picture breaks down, and the quenching of the SOC splitting occurs not because of the non-cubic crystal fields, but because of the covalency between neighboring Ir $d$ orbitals. 
%Especially it is shown that the face-sharing geometry between Ir octahedra is the origin of the large intermetallic covalency, replacing the local $j_{\rm eff}$ moment picture by a molecular-orbital (MO) description.
%
The main message of this work is illustrated in Fig.~\ref{fig:Fig0}, where the three representative local geometries consisting of IrO$_6$ octahedra --- corner-, edge-, and face-sharing structures --- are depicted. In terms of covalency, a critical difference between the three structures is the bond length between the nearest-neighboring Ir sites, which determines the strength of the Ir $d$-$d$ direct overlap integral $t_{dd}$~\cite{Streltsov2016}. While $t_{dd}$ tends to be smaller than the size of SOC ($\lambda_{\rm SO}$) for the corner- and edge-sharing geometries (Fig.~\ref{fig:Fig0}(a))\footnote{Several recent reports of pressure-induced Ir dimerizations in layered- and hyper-honeycomb iridates ~\cite{Veiga2017,Hermann2018} implies that, in edge-sharing geometries, $t_{dd}$ is almost comparable to $\lambda_{\rm SO}$, so that relatively small pressure of $<$5 GPa is enough to enhance $t_{dd}$ to break the $j_{\rm eff}$\,=\,1/2 states in these compounds.}, it can be stronger than $\lambda_{\rm SO}$ for the face-sharing structures because of the shorter Ir-Ir distance. In such cases, the neighboring Ir sites should form {\it molecular orbitals} (MO) as depicted in Fig.~\ref{fig:Fig0}(b). Therefore the $j_{\rm eff}$\,=\,1/2 local moment picture in the face-sharing geometry breaks down and the effects of SOC and Coulomb interactions should be considered based on the MO description. 

\begin{figure}
\includegraphics[width=0.30\textwidth]{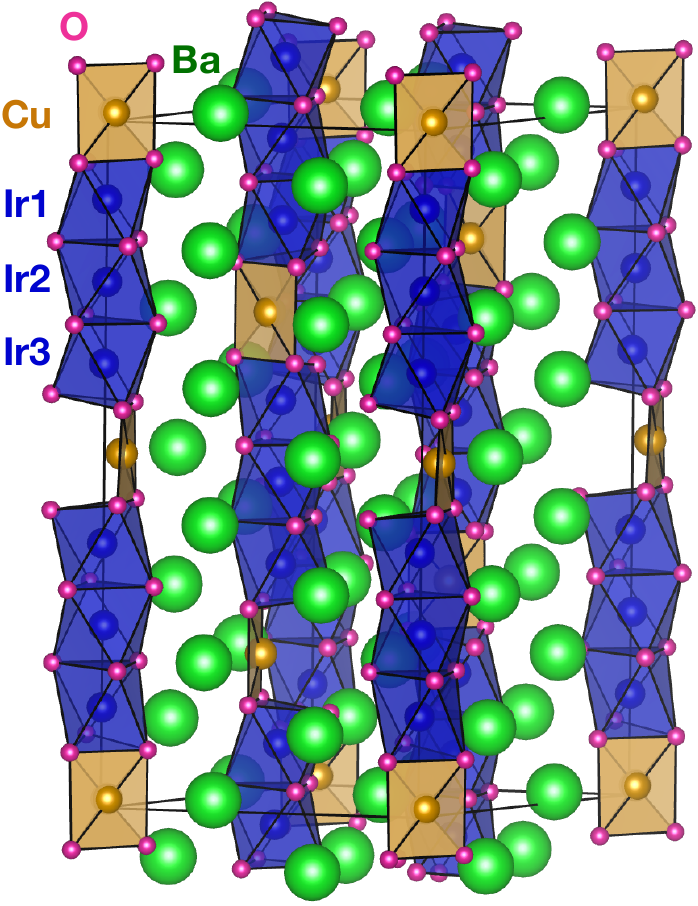}
\caption{\label{fig:FigA}Ba$_5$CuIr$_3$O$_{12}$ crystal structure employed for the \textit{ab-initio} calculations. Copper atoms are at the center of the prism face.}
\end{figure}

By combining Raman spectroscopy measurements and {\it ab-initio} theoretical analyses, we study a mixed 3$d$-5$d$ insulator Ba$_5$CuIr$_3$O$_{12}$ for which the $j_{\rm eff}$ = 1/2 approach breaks down~\cite{Blake1998,Blake1999}. In this compound, sequences of trigonal prismatic and octahedral transition metal sites run in chains parallel to the crystallographic z-axis, with Ba atoms located between the chains [Fig.~\ref{fig:FigA}]. \textit{Ab-initio} calculations and a tight-binding (TB) analysis yield a MO description of the electronic structure originating from the face-sharing geometry as depicted in Fig.~\ref{fig:Fig0}(b), and predict a three-peak structure in the joint density of states (JDOS). Raman scattering, a technique successfully used to study electronic excitations in iridate compounds~\cite{Yang2015},
%\cite{Yamanaka1988,Reznik1992,Liu1993,Chen1994PRL,Salamon1995,Salamon1996,Sacuto1997,Nemetschek1997,Tacon2006,Sala2014,Yang2015},
%\cite{CARDONA1988,THOMSEN1988,Blumberg1996,Nemetschek1997,Opel2000,Yang2015,Gretarsson2016,Glamazda2016},
%\cite{Liu1993,Salamon1995,Nemetschek1997,Salamon1996,Yang2015}, 
verifies this prediction. We observe one strong and sharp excitation at 0.58\,eV, and two weak features at 0.66 and 0.74\,eV. It should be noted that such a three-peak structure is not observed in systems with well-defined $j_{\rm eff}$\,=\,1/2 local moments~\cite{Yang2015}.

%The \textit{ab-initio} calculations are performed by the Vienna {\it ab-initio} Simulation Package (VASP) which employs the p

%the DFT+$U$ method~\cite{DFTU} with SOC included, based on the VASP package~\cite{VASP1,VASP2}. The details are provided in the Supplemental Material.

\begin{figure}
\includegraphics[width=0.48\textwidth]{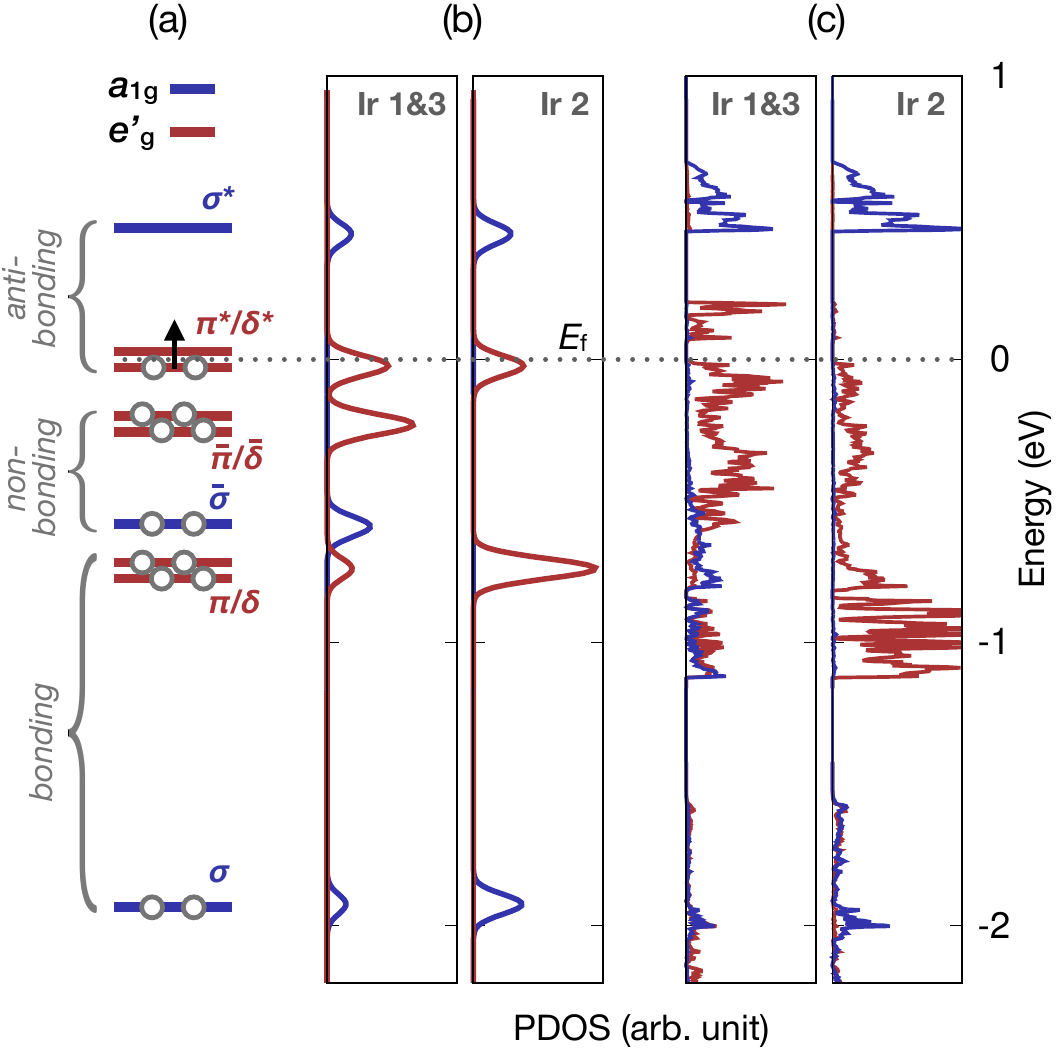}
\caption{\label{fig:FigB} (a) Energy level diagram showing the splitting of the Ir $t_{\rm 2g}$ states in an Ir trimer into MO states. Here $\sigma/\pi/\delta$, $\bar{\sigma}/\bar{\pi}/\bar{\delta}$, and $\sigma^*/\pi^*/\delta^*$ denote bonding, nonbonding, and antibonding states, respectively. Electrons in fully filled states are represented by circles, while magnetically active electrons are represented by arrows. (b,c) Projected DOS from the simple tight-bonding model for Ir trimers (b), and from \textit{ab-initio} calculations without SOC and magnetism (c). The color scheme for the orbital character is the same in (a-c).}
\end{figure}

We identify the structural motif for Ba$_5$CuIr$_3$O$_{12}$ to be the three face-sharing IrO$_6$ octahedra forming an Ir trimer as shown in Fig.~\ref{fig:Fig0}(b) and Fig.~\ref{fig:FigA}. If the intra-Ir-trimer hybridization dominates SOC, the three Ir $t_{\rm 2g}$ orbitals at the trimer sites split into nine MOs [Fig.~\ref{fig:FigB}(a)]. Among those, the atomic $a_{\rm 1g}$ singlet at each site gives rise to the strongest $\sigma$-type overlap between neighboring Ir sites; while the other $e'_{\rm g}$ doublets lead to weaker $\pi$- or $\delta$-like overlaps. Such scenario can be tested by constructing a simple TB model and comparing the results with those from DFT calculations. For the TB model, as a first-order approximation, we assume a three-fold symmetry along the $z$ direction and ignore the Cu-Ir hybridization. After this simplification, just four free parameters are left for the Ir trimer model where the parameters are tuned to fit the DFT DOS afterwards~\footnote{For the DFT calculations we employed the Vienna {\it ab-initio} Simulation Package (VASP)~\cite{VASP1,VASP2}. For the details of the TB model and DFT calculations please refer to Supplementary Material.}. Note that, because this is a test for the molecular orbital picture, SOC and the on-site Coulomb interactions are not considered at this stage.

Fig.~\ref{fig:FigB} presents the comparison between the TB model and the DFT results, where Fig.~\ref{fig:FigB}(a), (b), and (c) show the schematic TB energy diagram, TB DOS, and DFT DOS respectively. Both the $a_{\rm 1g}$- and $e'_{\rm g}$-derived MO states, depicted in blue and red in Fig.~\ref{fig:FigB}, show bonding ($\sigma/\pi/\delta$), nonbonding ($\bar{\sigma}/\bar{\pi}/\bar{\delta}$), and antibonding $\sigma^*/\pi^*/\delta^*$ characters [Fig.~\ref{fig:FigB}(a)]. Remarkably, the DOS from the simple 4-parameter-model agrees quite well with the DFT DOS; features of the DOS obtained from DFT calculations are consistent with those derived from the TB analysis, especially that the nonbonding $\bar{\sigma}/\bar{\pi}/\bar{\delta}$-MOs are located only at the two ends of the trimer [Fig.~\ref{fig:FigB}(b) and (c)]. The bonding-antibonding splitting between the $\sigma$- and $\sigma^*$-MO is 2.5\,eV, much larger than the strength of SOC (0.4\,eV)~\cite{Kim2008}. Therefore the intra-Ir-trimer hybridization indeed dominates SOC. 

\begin{figure}
\includegraphics[width=0.49\textwidth]{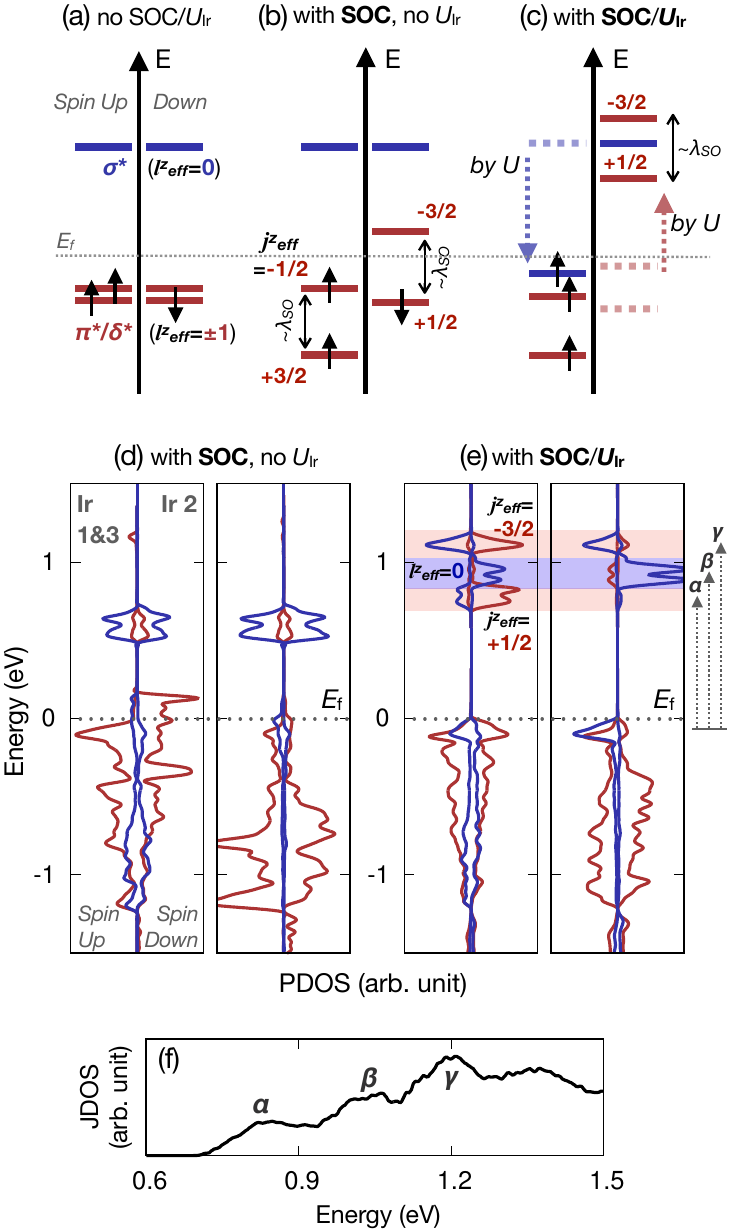}
\caption{\label{fig:FigC}(a-c) Schematic diagram showing the MO levels (a) with both SOC and $U_{\rm Ir}$ not included, (b) with SOC included but with no $U_{\rm Ir}$, and (c) both SOC and $U_{\rm Ir}$ included ($U_{\rm Ir}$ denoting $U$ at Ir sites). The $l^z_{\rm eff}$ and $j^z_{\rm eff}$ eigenvalues are given. (d-e) Projected DOS from DFT+SOC+U calculations for the $a_{\rm 1g}$ and $e'_{\rm g}$ states at Ir sites, where the $U$ values employed are ($U_{\rm Ir}, U_{\rm Cu}$) = (0, 6)eV and (2.8, 6)eV for (d) and (e), respectively ($U_{\rm Cu}$ denoting $U$ at Cu sites). (d) The joint DOS (JDOS) with ($U_{\rm Ir}, U_{\rm Cu}$) = (2.8, 6)eV, showing a three-peak structure ($\alpha$, $\beta$, and $\gamma$).}
\end{figure}

Since the size of MO splitting is large, we only need to consider the effect of SOC near the Fermi level. Fig.~\ref{fig:FigC} illustrates how SOC and the Coulomb interaction induce a spin polarization within the $\sigma^*/\pi^*/\delta^*$ MOs and in turn open a gap. First, because the $\pi^*/\delta^*$ MOs carry atomic orbital angular momenta $l_{\rm eff}^z$\,=\,$\pm$\,1, SOC functions as an orbital Zeeman field that splits the $\pi^*/\delta^*$ MOs [compare Fig.~\ref{fig:FigC}(a) and (b)]. Then the Coulomb interaction induces a `high-spin-like' configuration by pushing down the unoccupied $\sigma^*$ state with $l_{\rm eff}^z$\,=\,0 below the Fermi level in the majority spin channel and fully spin-polarizing the $\pi^*/\delta^*$ and $\sigma^*$ states as shown in Fig.~\ref{fig:FigC}(c). Fig.~\ref{fig:FigC}(d) and (e) show the projected DOS with $U_{\rm Ir}$ = 0 and 2.8 eV respectively ($U_{\rm Ir}$ denoting $U$ at Ir sites), where the $a_{\rm 1g}$/$e'_{\rm g}$-projected DOS from the DFT+$U$~\cite{DFTU} calculation with SOC included is plotted. Comparing Fig.~\ref{fig:FigC}(d) and (e) demonstrates the spin-polarizing effect of $U_{\rm Ir}$. As a result, a three-peak structure appears in the unoccupied sector as shown in Fig.~\ref{fig:FigC}(e). The three peaks, $\alpha$, $\beta$, and $\gamma$ in the JDOS from the DFT+$U$ result [Fig.~\ref{fig:FigC}(d)] originate from the transitions from the highest occupied MO state with $j_{\rm eff}^z$\,=\,0 to the unoccupied MO states with $j_{\rm eff}^z$ = $+1/2$, $0$, and $-3/2$, respectively. Note that the inclusion of $U_{\rm Ir}$ tends to recover the local atomic picture by mixing MO states, as shown in Fig.~\ref{fig:FigC}(c), where there is a small mixture among the $j_{\rm eff}^z$ = $+1/2$, $0$, and $-3/2$ MO states. This effect, however, does not qualitatively affects the above MO description. Note also that different values of $U_{\rm Ir}$ only changes the gap size while not affecting the three-peak structure as shown in the Supplementary Materials.

To confirm the predicted three-peak structure in the electronic excitation spectrum, we perform Raman-scattering measurements in a quasi-back-scattering geometry from the (001) crystallographic surface of Ba$_5$CuIr$_3$O$_{12}$ single crystal grown by flux method~[see Supplementary Materials for details of sample preparation and Raman scattering]. We use 476.2\,nm line from a Kr$^+$ ion laser for excitation. Incident light with $\sim$10\,mW power is focused to a 50$\times$100\,$\mu$m$^{2}$ spot.

\begin{figure}
\includegraphics[width=0.45\textwidth]{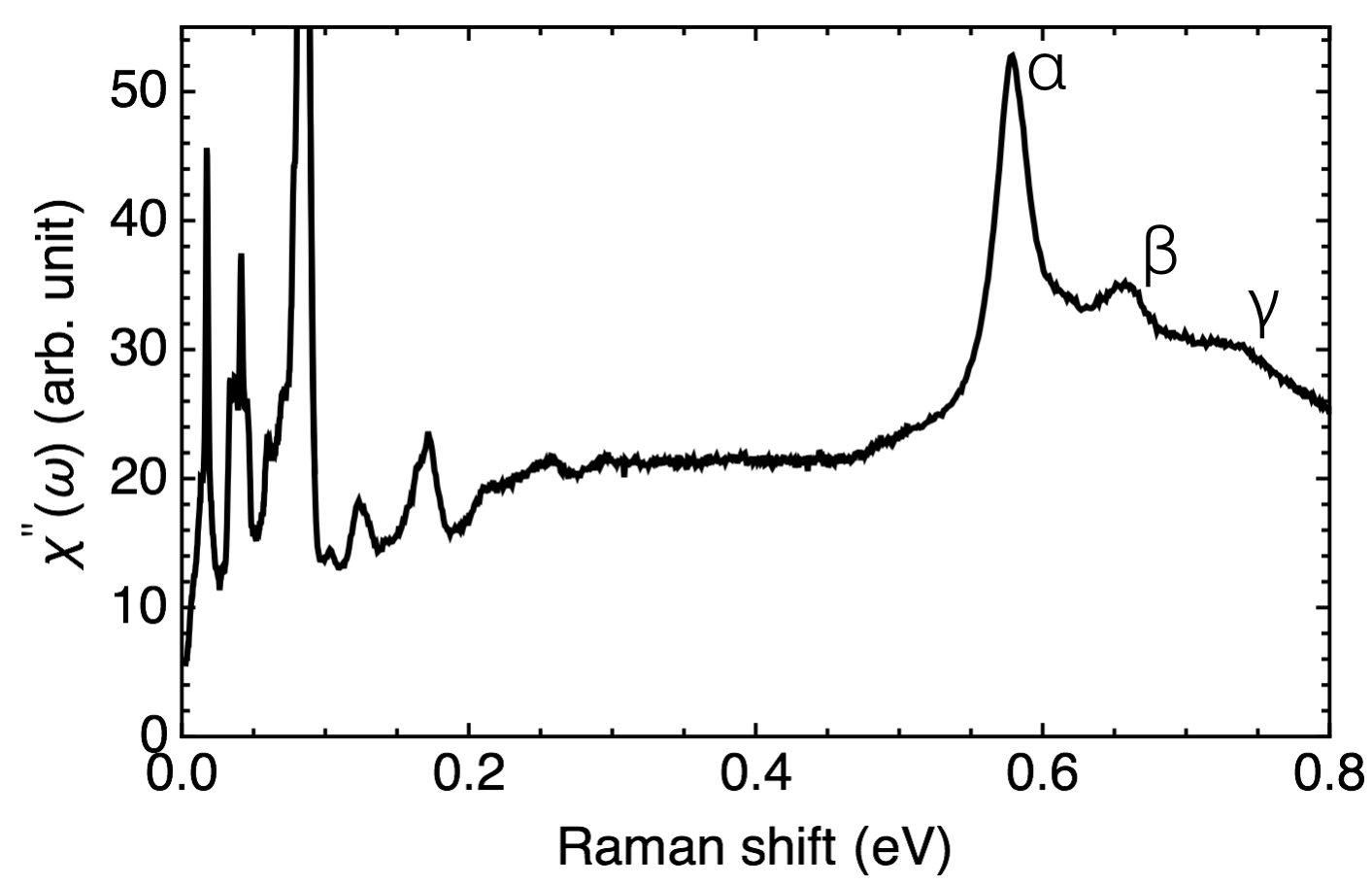}
\caption{\label{fig:Spectrum} Raman spectrum $\chi''(\omega)$ at 25\,K. Sharp features below 0.1\,eV are phonon modes; the two peaks at 0.13 and 0.17\,eV result from second-order phonon scattering while the broad feature at 0.24\,meV originates from third-order phonon scattering. The three high-energy electronic excitations at 0.58, 0.66 and 0.74\,eV are labeled by $\alpha$, $\beta$, and $\gamma$ respectively, corresponding to the labeling in Fig.~\ref{fig:FigC}(d).}
\end{figure}

Fig.~\ref{fig:Spectrum} shows the Raman spectrum measured at 25\,K. The sharp features at 17, 41 and 84\,meV are phonon modes~[see Supplementary Materials for the low-energy Raman spectrum]. The two peaks at 130 and 170\,meV result from second-order phonon scattering (41\,+\,84\,meV and 84\,+\,84\,meV, respectively). The broad feature centered at 240\,meV, weaker and broader than the second-order phonon scattering peaks, is attributed to third-order phonon scattering (84\,+\,84\,+\,84\,meV).

Importantly, three high-energy electronic excitations at 0.58, 0.66 and 0.74\,eV are resolved. The high-energy Raman spectrum exhibits the general trends of the JDOS from DFT+$U$. The high-energy gap in the Raman spectrum is about 0.55\,eV, while it is 0.70\,eV in the JDOS. The three peaks in the Raman spectrum are evenly spaced, with a separation of 0.08\,eV; those in the JDOS are also evenly spaced but with a 0.2\,eV separation. One possible reason for the smaller splitting in experimental spectrum compared to the DFT+$U$ JDOS peak splitting can be a stronger mixing between the $l_{\rm eff}^z$\,=\,$\pm$\,1 and 0 antibonding MO states near the Fermi level in the real system. This mixing reduces the expectation value of SOC energy and in turn decreases the separation. Another thing to mention is, while in the JDOS all the three peaks have similar spectral weight, in the Raman spectrum the 0.58\,eV peak is much stronger than the other two. This could be attributed to the matrix element effect.

The three-peak structure in the DFT+$U$ JDOS and in the Raman measurement is a distinct feature indicating deviation from the $j_{\rm eff}$\,=\,1/2 picture. Simple $j_{\rm eff}$\,=\,1/2 picture predicts up to two high-energy transition peaks, because non-cubic crystal fields just induce splitting of the fully occupied $j_{\rm eff}$\,=\,3/2 quartet~\cite{Kim2008,Gretarsson2013}. On the contrary, our three-peak structure in Ba$_5$CuIr$_3$O$_{12}$ comes from the strong Ir-Ir hybridization in the face-sharing IrO$_6$ octahedral geometry and the resulting formation of MOs. We note that in other compounds with similar face-sharing geometries with alternating 3$d$ and 5$d$ transition metal ions such as Sr$_3$NiIrO$_6$~\cite{Lefrancois2016} or Sr$_3$CuIrO$_6$~\cite{Liu2012}, two-peak structures are observed in their $j_{\rm eff}$-excitation spectra~\cite{Gretarsson2013}. This implies that, although the size of non-cubic distortions is large in these compounds, still the local $j_{\rm eff}$\,=\,1/2 moment picture remains effective because of the reduced covalency between the 3$d$ and 5$d$ orbitals as suggested in a recent {\it ab-initio} study for Sr$_3$NiIrO$_6$~\cite{Birol2018}. We also comment that there is another theoretical study on BaIrO$_3$~\cite{Ju2013}, consisting of the same face-sharing Ir$_3$O$_{12}$ octahedral trimers like Ba$_5$CuIr$_3$O$_{12}$. A similar three-peak structure in the upper Hubbard band (UHB) was reported therein, but it was speculated that the UHB states still retain the $j_{\rm eff}$\,=\,1/2 character, which seems to require a more rigorous justification. 

As for possible magnetism in this compound; from the projected DOS plotted in Fig.~\ref{fig:FigC}(b) and (c), we identify three magnetically active states; a $\sigma^*$ state with mostly $a_{\rm 1g}$ ($l^z_{\rm eff}$\,=\,0) character mainly located at Ir 2 site, and two $\pi^*/\delta^*$ states with the $e'_{\rm g}$ character ($l^z_{\rm eff}$\,=\,$\pm$\,1) at Ir 1 and 3 sites. The strong SOC within the Ir $d$-orbital then behaves as a single-ion anisotropy to the electron spins filling the $l^z_{\rm eff}$\,=\,$\pm$\,1 MO states, locking the spins parallel to the $z$-direction. Spin in the $l^z_{\rm eff}$\,=\,0 MO state, on the other hand, has little single-ion anisotropy due to the vanishing orbital angular momentum. The spin moments at Cu sites are also isotropic, so this compound should have three different kinds of spin moments: isotropic Cu spins, isotropic Ir spins at Ir 2 sites, and anisotropic Ir spins at Ir 1 and 3 sites locked along the $z$-direction. Because all the Ir spins are occupying the MO states, rather than behaving as the $j_{\rm eff}$\,=\,1/2 local moments, they may show distinct low-energy magnetic properties compared to previously known magnetic iridate compounds. For future studies, interesting questions about the outcome of MO formation can be posed, for example on the form of exchange interactions and the spectrum of low-energy magnetic excitations.

Our study on the face-sharing iridate Ba$_5$CuIr$_3$O$_{12}$ demonstrates the breakdown of the SOC-based $j_{\rm eff}$\,=\,1/2 physics, and reveals the MO nature of the electronic structure originating from the strong intermetallic $d$-$d$ direct overlap. A similar scenario, leading to the formation of benzene-ring-shaped quasi-molecular orbitals (QMO) driven by $\pi$-like $d$-$p$ overlap, was suggested for Na$_2$IrO$_3$~\cite{Mazin2012}. In Na$_2$IrO$_3$, the $d$-$p$ overlap preserves the three-fold symmetry of the Ir $t_{\rm 2g}$ orbitals, hence the inclusion of SOC and $U$ induces a crossover from the delocalized QMO to the local $j_{\rm eff}$\,=\,1/2 moment picture~\cite{Foyevtsova2013}. In Ba$_5$CuIr$_3$O$_{12}$, on the contrary, the direct overlap $t_\sigma$ is not only huge but also explicitly breaks the degeneracy of the Ir $t_{\rm 2g}$ orbitals, resulting in a completely different MO description. Overall, this work suggest a peculiar relation between the crystal structure and the nature of electronic degree of freedom in 5$d$ iridate and other transition metal compounds, which can be useful in search of novel correlated materials.

\begin{acknowledgments}
M.Y. thanks H.-H. Kung for helpful comments on the early version of the manuscript. Crystal growth (C.-J. W.) was supported by the National Research Foundation of Korea, Ministry of Science and ICT (No. 2016K1A4A4A01922028). The DFT study (H.-S.K., D.V., K.H.) and and the crystal characterization (J.-W.K., S.-W.C.) were supported by NSF DMREF DMR-1629059. The spectroscopic work (M.Y., G.B.) was supported by NSF Grant No. DMR-1709161.

M.Y. performed the spectroscopic study and H.-S.K. did the \textit{ab-initio} calculations. These two authors contributed equally to this work.

\end{acknowledgments}

\bibliography{BCIO.bib}

\begin{thebibliography}{38}%
\makeatletter
\providecommand \@ifxundefined [1]{%
 \@ifx{#1\undefined}
}%
\providecommand \@ifnum [1]{%
 \ifnum #1\expandafter \@firstoftwo
 \else \expandafter \@secondoftwo
 \fi
}%
\providecommand \@ifx [1]{%
 \ifx #1\expandafter \@firstoftwo
 \else \expandafter \@secondoftwo
 \fi
}%
\providecommand \natexlab [1]{#1}%
\providecommand \enquote  [1]{``#1''}%
\providecommand \bibnamefont  [1]{#1}%
\providecommand \bibfnamefont [1]{#1}%
\providecommand \citenamefont [1]{#1}%
\providecommand \href@noop [0]{\@secondoftwo}%
\providecommand \href [0]{\begingroup \@sanitize@url \@href}%
\providecommand \@href[1]{\@@startlink{#1}\@@href}%
\providecommand \@@href[1]{\endgroup#1\@@endlink}%
\providecommand \@sanitize@url [0]{\catcode `\\12\catcode `\$12\catcode
  `\&12\catcode `\#12\catcode `\^12\catcode `\_12\catcode `\%12\relax}%
\providecommand \@@startlink[1]{}%
\providecommand \@@endlink[0]{}%
\providecommand \url  [0]{\begingroup\@sanitize@url \@url }%
\providecommand \@url [1]{\endgroup\@href {#1}{\urlprefix }}%
\providecommand \urlprefix  [0]{URL }%
\providecommand \Eprint [0]{\href }%
\providecommand \doibase [0]{http://dx.doi.org/}%
\providecommand \selectlanguage [0]{\@gobble}%
\providecommand \bibinfo  [0]{\@secondoftwo}%
\providecommand \bibfield  [0]{\@secondoftwo}%
\providecommand \translation [1]{[#1]}%
\providecommand \BibitemOpen [0]{}%
\providecommand \bibitemStop [0]{}%
\providecommand \bibitemNoStop [0]{.\EOS\space}%
\providecommand \EOS [0]{\spacefactor3000\relax}%
\providecommand \BibitemShut  [1]{\csname bibitem#1\endcsname}%
\let\auto@bib@innerbib\@empty
%</preamble>
\bibitem [{\citenamefont {Imada}\ \emph {et~al.}(1998)\citenamefont {Imada},
  \citenamefont {Fujimori},\ and\ \citenamefont {Tokura}}]{Imada1998}%
  \BibitemOpen
  \bibfield  {author} {\bibinfo {author} {\bibfnamefont {M.}~\bibnamefont
  {Imada}}, \bibinfo {author} {\bibfnamefont {A.}~\bibnamefont {Fujimori}}, \
  and\ \bibinfo {author} {\bibfnamefont {Y.}~\bibnamefont {Tokura}},\ }\href
  {\doibase 10.1103/RevModPhys.70.1039} {\bibfield  {journal} {\bibinfo
  {journal} {Rev. Mod. Phys.}\ }\textbf {\bibinfo {volume} {70}},\ \bibinfo
  {pages} {1039} (\bibinfo {year} {1998})}\BibitemShut {NoStop}%
\bibitem [{\citenamefont {Khomskii}(2014)}]{Khomskii2014}%
  \BibitemOpen
  \bibfield  {author} {\bibinfo {author} {\bibfnamefont {D.~I.}\ \bibnamefont
  {Khomskii}},\ }\href@noop {} {\emph {\bibinfo {title} {Transition Metal
  Compounds}}}\ (\bibinfo  {publisher} {Cambridge University Press},\ \bibinfo
  {address} {Cambridge},\ \bibinfo {year} {2014})\BibitemShut {NoStop}%
\bibitem [{\citenamefont {Cox}(1992)}]{Cox1992}%
  \BibitemOpen
  \bibfield  {author} {\bibinfo {author} {\bibfnamefont {P.~A.}\ \bibnamefont
  {Cox}},\ }\enquote {\bibinfo {title} {Metallic oxides},}\ in\ \href@noop {}
  {\emph {\bibinfo {booktitle} {Transition Metal Oxides: An Introduction to
  Their Electronic Structure and Properties}}}\ (\bibinfo  {publisher} {Oxford
  University Press},\ \bibinfo {address} {Oxford},\ \bibinfo {year} {1992})\
  pp.\ \bibinfo {pages} {204--276}\BibitemShut {NoStop}%
\bibitem [{\citenamefont {Damascelli}\ \emph {et~al.}(2000)\citenamefont
  {Damascelli}, \citenamefont {Lu}, \citenamefont {Shen}, \citenamefont
  {Armitage}, \citenamefont {Ronning}, \citenamefont {Feng}, \citenamefont
  {Kim}, \citenamefont {Shen}, \citenamefont {Kimura}, \citenamefont {Tokura},
  \citenamefont {Mao},\ and\ \citenamefont {Maeno}}]{Damascelli2000}%
  \BibitemOpen
  \bibfield  {author} {\bibinfo {author} {\bibfnamefont {A.}~\bibnamefont
  {Damascelli}}, \bibinfo {author} {\bibfnamefont {D.~H.}\ \bibnamefont {Lu}},
  \bibinfo {author} {\bibfnamefont {K.~M.}\ \bibnamefont {Shen}}, \bibinfo
  {author} {\bibfnamefont {N.~P.}\ \bibnamefont {Armitage}}, \bibinfo {author}
  {\bibfnamefont {F.}~\bibnamefont {Ronning}}, \bibinfo {author} {\bibfnamefont
  {D.~L.}\ \bibnamefont {Feng}}, \bibinfo {author} {\bibfnamefont
  {C.}~\bibnamefont {Kim}}, \bibinfo {author} {\bibfnamefont {Z.-X.}\
  \bibnamefont {Shen}}, \bibinfo {author} {\bibfnamefont {T.}~\bibnamefont
  {Kimura}}, \bibinfo {author} {\bibfnamefont {Y.}~\bibnamefont {Tokura}},
  \bibinfo {author} {\bibfnamefont {Z.~Q.}\ \bibnamefont {Mao}}, \ and\
  \bibinfo {author} {\bibfnamefont {Y.}~\bibnamefont {Maeno}},\ }\href
  {\doibase 10.1103/PhysRevLett.85.5194} {\bibfield  {journal} {\bibinfo
  {journal} {Phys. Rev. Lett.}\ }\textbf {\bibinfo {volume} {85}},\ \bibinfo
  {pages} {5194} (\bibinfo {year} {2000})}\BibitemShut {NoStop}%
\bibitem [{\citenamefont {Baumberger}\ \emph {et~al.}(2006)\citenamefont
  {Baumberger}, \citenamefont {Ingle}, \citenamefont {Meevasana}, \citenamefont
  {Shen}, \citenamefont {Lu}, \citenamefont {Perry}, \citenamefont {Mackenzie},
  \citenamefont {Hussain}, \citenamefont {Singh},\ and\ \citenamefont
  {Shen}}]{Baumberger2006}%
  \BibitemOpen
  \bibfield  {author} {\bibinfo {author} {\bibfnamefont {F.}~\bibnamefont
  {Baumberger}}, \bibinfo {author} {\bibfnamefont {N.~J.~C.}\ \bibnamefont
  {Ingle}}, \bibinfo {author} {\bibfnamefont {W.}~\bibnamefont {Meevasana}},
  \bibinfo {author} {\bibfnamefont {K.~M.}\ \bibnamefont {Shen}}, \bibinfo
  {author} {\bibfnamefont {D.~H.}\ \bibnamefont {Lu}}, \bibinfo {author}
  {\bibfnamefont {R.~S.}\ \bibnamefont {Perry}}, \bibinfo {author}
  {\bibfnamefont {A.~P.}\ \bibnamefont {Mackenzie}}, \bibinfo {author}
  {\bibfnamefont {Z.}~\bibnamefont {Hussain}}, \bibinfo {author} {\bibfnamefont
  {D.~J.}\ \bibnamefont {Singh}}, \ and\ \bibinfo {author} {\bibfnamefont
  {Z.-X.}\ \bibnamefont {Shen}},\ }\href {\doibase
  10.1103/PhysRevLett.96.246402} {\bibfield  {journal} {\bibinfo  {journal}
  {Phys. Rev. Lett.}\ }\textbf {\bibinfo {volume} {96}},\ \bibinfo {pages}
  {246402} (\bibinfo {year} {2006})}\BibitemShut {NoStop}%
\bibitem [{\citenamefont {Crawford}\ \emph {et~al.}(1994)\citenamefont
  {Crawford}, \citenamefont {Subramanian}, \citenamefont {Harlow},
  \citenamefont {Fernandez-Baca}, \citenamefont {Wang},\ and\ \citenamefont
  {Johnston}}]{Crawford1994}%
  \BibitemOpen
  \bibfield  {author} {\bibinfo {author} {\bibfnamefont {M.~K.}\ \bibnamefont
  {Crawford}}, \bibinfo {author} {\bibfnamefont {M.~A.}\ \bibnamefont
  {Subramanian}}, \bibinfo {author} {\bibfnamefont {R.~L.}\ \bibnamefont
  {Harlow}}, \bibinfo {author} {\bibfnamefont {J.~A.}\ \bibnamefont
  {Fernandez-Baca}}, \bibinfo {author} {\bibfnamefont {Z.~R.}\ \bibnamefont
  {Wang}}, \ and\ \bibinfo {author} {\bibfnamefont {D.~C.}\ \bibnamefont
  {Johnston}},\ }\href {\doibase 10.1103/PhysRevB.49.9198} {\bibfield
  {journal} {\bibinfo  {journal} {Phys. Rev. B}\ }\textbf {\bibinfo {volume}
  {49}},\ \bibinfo {pages} {9198} (\bibinfo {year} {1994})}\BibitemShut
  {NoStop}%
\bibitem [{\citenamefont {Okabe}\ \emph {et~al.}(2011)\citenamefont {Okabe},
  \citenamefont {Isobe}, \citenamefont {Takayama-Muromachi}, \citenamefont
  {Koda}, \citenamefont {Takeshita}, \citenamefont {Hiraishi}, \citenamefont
  {Miyazaki}, \citenamefont {Kadono}, \citenamefont {Miyake},\ and\
  \citenamefont {Akimitsu}}]{Okabe2011}%
  \BibitemOpen
  \bibfield  {author} {\bibinfo {author} {\bibfnamefont {H.}~\bibnamefont
  {Okabe}}, \bibinfo {author} {\bibfnamefont {M.}~\bibnamefont {Isobe}},
  \bibinfo {author} {\bibfnamefont {E.}~\bibnamefont {Takayama-Muromachi}},
  \bibinfo {author} {\bibfnamefont {A.}~\bibnamefont {Koda}}, \bibinfo {author}
  {\bibfnamefont {S.}~\bibnamefont {Takeshita}}, \bibinfo {author}
  {\bibfnamefont {M.}~\bibnamefont {Hiraishi}}, \bibinfo {author}
  {\bibfnamefont {M.}~\bibnamefont {Miyazaki}}, \bibinfo {author}
  {\bibfnamefont {R.}~\bibnamefont {Kadono}}, \bibinfo {author} {\bibfnamefont
  {Y.}~\bibnamefont {Miyake}}, \ and\ \bibinfo {author} {\bibfnamefont
  {J.}~\bibnamefont {Akimitsu}},\ }\href {\doibase 10.1103/PhysRevB.83.155118}
  {\bibfield  {journal} {\bibinfo  {journal} {Phys. Rev. B}\ }\textbf {\bibinfo
  {volume} {83}},\ \bibinfo {pages} {155118} (\bibinfo {year}
  {2011})}\BibitemShut {NoStop}%
\bibitem [{\citenamefont {Singh}\ and\ \citenamefont
  {Gegenwart}(2010)}]{Singh2010}%
  \BibitemOpen
  \bibfield  {author} {\bibinfo {author} {\bibfnamefont {Y.}~\bibnamefont
  {Singh}}\ and\ \bibinfo {author} {\bibfnamefont {P.}~\bibnamefont
  {Gegenwart}},\ }\href {\doibase 10.1103/PhysRevB.82.064412} {\bibfield
  {journal} {\bibinfo  {journal} {Phys. Rev. B}\ }\textbf {\bibinfo {volume}
  {82}},\ \bibinfo {pages} {064412} (\bibinfo {year} {2010})}\BibitemShut
  {NoStop}%
\bibitem [{\citenamefont {Ohgushi}\ \emph {et~al.}(2013)\citenamefont
  {Ohgushi}, \citenamefont {Yamaura}, \citenamefont {Ohsumi}, \citenamefont
  {Sugimoto}, \citenamefont {Takeshita}, \citenamefont {Tokuda}, \citenamefont
  {Takagi}, \citenamefont {Takata},\ and\ \citenamefont {Arima}}]{Ohgushi2013}%
  \BibitemOpen
  \bibfield  {author} {\bibinfo {author} {\bibfnamefont {K.}~\bibnamefont
  {Ohgushi}}, \bibinfo {author} {\bibfnamefont {J.-i.}\ \bibnamefont
  {Yamaura}}, \bibinfo {author} {\bibfnamefont {H.}~\bibnamefont {Ohsumi}},
  \bibinfo {author} {\bibfnamefont {K.}~\bibnamefont {Sugimoto}}, \bibinfo
  {author} {\bibfnamefont {S.}~\bibnamefont {Takeshita}}, \bibinfo {author}
  {\bibfnamefont {A.}~\bibnamefont {Tokuda}}, \bibinfo {author} {\bibfnamefont
  {H.}~\bibnamefont {Takagi}}, \bibinfo {author} {\bibfnamefont
  {M.}~\bibnamefont {Takata}}, \ and\ \bibinfo {author} {\bibfnamefont {T.-h.}\
  \bibnamefont {Arima}},\ }\href {\doibase 10.1103/PhysRevLett.110.217212}
  {\bibfield  {journal} {\bibinfo  {journal} {Phys. Rev. Lett.}\ }\textbf
  {\bibinfo {volume} {110}},\ \bibinfo {pages} {217212} (\bibinfo {year}
  {2013})}\BibitemShut {NoStop}%
\bibitem [{\citenamefont {Kim}\ \emph {et~al.}(2008)\citenamefont {Kim},
  \citenamefont {Jin}, \citenamefont {Moon}, \citenamefont {Kim}, \citenamefont
  {Park}, \citenamefont {Leem}, \citenamefont {Yu}, \citenamefont {Noh},
  \citenamefont {Kim}, \citenamefont {Oh}, \citenamefont {Park}, \citenamefont
  {Durairaj}, \citenamefont {Cao},\ and\ \citenamefont {Rotenberg}}]{Kim2008}%
  \BibitemOpen
  \bibfield  {author} {\bibinfo {author} {\bibfnamefont {B.~J.}\ \bibnamefont
  {Kim}}, \bibinfo {author} {\bibfnamefont {H.}~\bibnamefont {Jin}}, \bibinfo
  {author} {\bibfnamefont {S.~J.}\ \bibnamefont {Moon}}, \bibinfo {author}
  {\bibfnamefont {J.-Y.}\ \bibnamefont {Kim}}, \bibinfo {author} {\bibfnamefont
  {B.-G.}\ \bibnamefont {Park}}, \bibinfo {author} {\bibfnamefont {C.~S.}\
  \bibnamefont {Leem}}, \bibinfo {author} {\bibfnamefont {J.}~\bibnamefont
  {Yu}}, \bibinfo {author} {\bibfnamefont {T.~W.}\ \bibnamefont {Noh}},
  \bibinfo {author} {\bibfnamefont {C.}~\bibnamefont {Kim}}, \bibinfo {author}
  {\bibfnamefont {S.-J.}\ \bibnamefont {Oh}}, \bibinfo {author} {\bibfnamefont
  {J.-H.}\ \bibnamefont {Park}}, \bibinfo {author} {\bibfnamefont
  {V.}~\bibnamefont {Durairaj}}, \bibinfo {author} {\bibfnamefont
  {G.}~\bibnamefont {Cao}}, \ and\ \bibinfo {author} {\bibfnamefont
  {E.}~\bibnamefont {Rotenberg}},\ }\href {\doibase
  10.1103/PhysRevLett.101.076402} {\bibfield  {journal} {\bibinfo  {journal}
  {Phys. Rev. Lett.}\ }\textbf {\bibinfo {volume} {101}},\ \bibinfo {pages}
  {076402} (\bibinfo {year} {2008})}\BibitemShut {NoStop}%
\bibitem [{\citenamefont {Kim}\ \emph {et~al.}(2009)\citenamefont {Kim},
  \citenamefont {Ohsumi}, \citenamefont {Komesu}, \citenamefont {Sakai},
  \citenamefont {Morita}, \citenamefont {Takagi},\ and\ \citenamefont
  {Arima}}]{Kim2009}%
  \BibitemOpen
  \bibfield  {author} {\bibinfo {author} {\bibfnamefont {B.~J.}\ \bibnamefont
  {Kim}}, \bibinfo {author} {\bibfnamefont {H.}~\bibnamefont {Ohsumi}},
  \bibinfo {author} {\bibfnamefont {T.}~\bibnamefont {Komesu}}, \bibinfo
  {author} {\bibfnamefont {S.}~\bibnamefont {Sakai}}, \bibinfo {author}
  {\bibfnamefont {T.}~\bibnamefont {Morita}}, \bibinfo {author} {\bibfnamefont
  {H.}~\bibnamefont {Takagi}}, \ and\ \bibinfo {author} {\bibfnamefont
  {T.}~\bibnamefont {Arima}},\ }\href {\doibase 10.1126/science.1167106}
  {\bibfield  {journal} {\bibinfo  {journal} {Science}\ }\textbf {\bibinfo
  {volume} {323}},\ \bibinfo {pages} {1329} (\bibinfo {year}
  {2009})}\BibitemShut {NoStop}%
\bibitem [{\citenamefont {Kim}\ \emph {et~al.}(2014)\citenamefont {Kim},
  \citenamefont {Krupin}, \citenamefont {Denlinger}, \citenamefont {Bostwick},
  \citenamefont {Rotenberg}, \citenamefont {Zhao}, \citenamefont {Mitchell},
  \citenamefont {Allen},\ and\ \citenamefont {Kim}}]{Kim2014}%
  \BibitemOpen
  \bibfield  {author} {\bibinfo {author} {\bibfnamefont {Y.~K.}\ \bibnamefont
  {Kim}}, \bibinfo {author} {\bibfnamefont {O.}~\bibnamefont {Krupin}},
  \bibinfo {author} {\bibfnamefont {J.~D.}\ \bibnamefont {Denlinger}}, \bibinfo
  {author} {\bibfnamefont {A.}~\bibnamefont {Bostwick}}, \bibinfo {author}
  {\bibfnamefont {E.}~\bibnamefont {Rotenberg}}, \bibinfo {author}
  {\bibfnamefont {Q.}~\bibnamefont {Zhao}}, \bibinfo {author} {\bibfnamefont
  {J.~F.}\ \bibnamefont {Mitchell}}, \bibinfo {author} {\bibfnamefont {J.~W.}\
  \bibnamefont {Allen}}, \ and\ \bibinfo {author} {\bibfnamefont {B.~J.}\
  \bibnamefont {Kim}},\ }\href {\doibase 10.1126/science.1251151} {\bibfield
  {journal} {\bibinfo  {journal} {Science}\ }\textbf {\bibinfo {volume}
  {345}},\ \bibinfo {pages} {187} (\bibinfo {year} {2014})}\BibitemShut
  {NoStop}%
\bibitem [{\citenamefont {Kim}\ \emph {et~al.}(2015{\natexlab{a}})\citenamefont
  {Kim}, \citenamefont {Sung}, \citenamefont {Denlinger},\ and\ \citenamefont
  {Kim}}]{Kim2015}%
  \BibitemOpen
  \bibfield  {author} {\bibinfo {author} {\bibfnamefont {Y.~K.}\ \bibnamefont
  {Kim}}, \bibinfo {author} {\bibfnamefont {N.~H.}\ \bibnamefont {Sung}},
  \bibinfo {author} {\bibfnamefont {J.~D.}\ \bibnamefont {Denlinger}}, \ and\
  \bibinfo {author} {\bibfnamefont {B.~J.}\ \bibnamefont {Kim}},\ }\href
  {\doibase 10.1038/nphys3503
  https://www.nature.com/articles/nphys3503#supplementary-information}
  {\bibfield  {journal} {\bibinfo  {journal} {Nature Physics}\ }\textbf
  {\bibinfo {volume} {12}},\ \bibinfo {pages} {37} (\bibinfo {year}
  {2015}{\natexlab{a}})}\BibitemShut {NoStop}%
\bibitem [{\citenamefont {Jackeli}\ and\ \citenamefont
  {Khaliullin}(2009)}]{Jackeli2009}%
  \BibitemOpen
  \bibfield  {author} {\bibinfo {author} {\bibfnamefont {G.}~\bibnamefont
  {Jackeli}}\ and\ \bibinfo {author} {\bibfnamefont {G.}~\bibnamefont
  {Khaliullin}},\ }\href {\doibase 10.1103/PhysRevLett.102.017205} {\bibfield
  {journal} {\bibinfo  {journal} {Phys. Rev. Lett.}\ }\textbf {\bibinfo
  {volume} {102}},\ \bibinfo {pages} {017205} (\bibinfo {year}
  {2009})}\BibitemShut {NoStop}%
\bibitem [{\citenamefont {Witczak-Krempa}\ \emph {et~al.}(2014)\citenamefont
  {Witczak-Krempa}, \citenamefont {Chen}, \citenamefont {Kim},\ and\
  \citenamefont {Balents}}]{Witczak-Krempa2014}%
  \BibitemOpen
  \bibfield  {author} {\bibinfo {author} {\bibfnamefont {W.}~\bibnamefont
  {Witczak-Krempa}}, \bibinfo {author} {\bibfnamefont {G.}~\bibnamefont
  {Chen}}, \bibinfo {author} {\bibfnamefont {Y.~B.}\ \bibnamefont {Kim}}, \
  and\ \bibinfo {author} {\bibfnamefont {L.}~\bibnamefont {Balents}},\ }\href
  {https://doi.org/10.1146/annurev-conmatphys-020911-125138} {\bibfield
  {journal} {\bibinfo  {journal} {Annu. Rev. Condens. Matter Phys.}\ }\textbf
  {\bibinfo {volume} {5}},\ \bibinfo {pages} {57} (\bibinfo {year}
  {2014})}\BibitemShut {NoStop}%
\bibitem [{\citenamefont {Rau}\ \emph {et~al.}(2016)\citenamefont {Rau},
  \citenamefont {Lee},\ and\ \citenamefont {Kee}}]{Rau2016}%
  \BibitemOpen
  \bibfield  {author} {\bibinfo {author} {\bibfnamefont {J.~G.}\ \bibnamefont
  {Rau}}, \bibinfo {author} {\bibfnamefont {E.~K.-H.}\ \bibnamefont {Lee}}, \
  and\ \bibinfo {author} {\bibfnamefont {H.-Y.}\ \bibnamefont {Kee}},\ }\href
  {\doibase 10.1146/annurev-conmatphys-031115-011319} {\bibfield  {journal}
  {\bibinfo  {journal} {Annu. Rev. Condens. Matter Phys.}\ }\textbf {\bibinfo
  {volume} {7}},\ \bibinfo {pages} {195} (\bibinfo {year} {2016})}\BibitemShut
  {NoStop}%
\bibitem [{\citenamefont {Zhang}\ \emph {et~al.}(2013)\citenamefont {Zhang},
  \citenamefont {Haule},\ and\ \citenamefont {Vanderbilt}}]{Zhang2013}%
  \BibitemOpen
  \bibfield  {author} {\bibinfo {author} {\bibfnamefont {H.}~\bibnamefont
  {Zhang}}, \bibinfo {author} {\bibfnamefont {K.}~\bibnamefont {Haule}}, \ and\
  \bibinfo {author} {\bibfnamefont {D.}~\bibnamefont {Vanderbilt}},\ }\href
  {\doibase 10.1103/PhysRevLett.111.246402} {\bibfield  {journal} {\bibinfo
  {journal} {Phys. Rev. Lett.}\ }\textbf {\bibinfo {volume} {111}},\ \bibinfo
  {pages} {246402} (\bibinfo {year} {2013})}\BibitemShut {NoStop}%
\bibitem [{\citenamefont {Gretarsson}\ \emph {et~al.}(2013)\citenamefont
  {Gretarsson}, \citenamefont {Clancy}, \citenamefont {Liu}, \citenamefont
  {Hill}, \citenamefont {Bozin}, \citenamefont {Singh}, \citenamefont {Manni},
  \citenamefont {Gegenwart}, \citenamefont {Kim}, \citenamefont {Said},
  \citenamefont {Casa}, \citenamefont {Gog}, \citenamefont {Upton},
  \citenamefont {Kim}, \citenamefont {Yu}, \citenamefont {Katukuri},
  \citenamefont {Hozoi}, \citenamefont {van~den Brink},\ and\ \citenamefont
  {Kim}}]{Gretarsson2013}%
  \BibitemOpen
  \bibfield  {author} {\bibinfo {author} {\bibfnamefont {H.}~\bibnamefont
  {Gretarsson}}, \bibinfo {author} {\bibfnamefont {J.~P.}\ \bibnamefont
  {Clancy}}, \bibinfo {author} {\bibfnamefont {X.}~\bibnamefont {Liu}},
  \bibinfo {author} {\bibfnamefont {J.~P.}\ \bibnamefont {Hill}}, \bibinfo
  {author} {\bibfnamefont {E.}~\bibnamefont {Bozin}}, \bibinfo {author}
  {\bibfnamefont {Y.}~\bibnamefont {Singh}}, \bibinfo {author} {\bibfnamefont
  {S.}~\bibnamefont {Manni}}, \bibinfo {author} {\bibfnamefont
  {P.}~\bibnamefont {Gegenwart}}, \bibinfo {author} {\bibfnamefont
  {J.}~\bibnamefont {Kim}}, \bibinfo {author} {\bibfnamefont {A.~H.}\
  \bibnamefont {Said}}, \bibinfo {author} {\bibfnamefont {D.}~\bibnamefont
  {Casa}}, \bibinfo {author} {\bibfnamefont {T.}~\bibnamefont {Gog}}, \bibinfo
  {author} {\bibfnamefont {M.~H.}\ \bibnamefont {Upton}}, \bibinfo {author}
  {\bibfnamefont {H.-S.}\ \bibnamefont {Kim}}, \bibinfo {author} {\bibfnamefont
  {J.}~\bibnamefont {Yu}}, \bibinfo {author} {\bibfnamefont {V.~M.}\
  \bibnamefont {Katukuri}}, \bibinfo {author} {\bibfnamefont {L.}~\bibnamefont
  {Hozoi}}, \bibinfo {author} {\bibfnamefont {J.}~\bibnamefont {van~den
  Brink}}, \ and\ \bibinfo {author} {\bibfnamefont {Y.-J.}\ \bibnamefont
  {Kim}},\ }\href {\doibase 10.1103/PhysRevLett.110.076402} {\bibfield
  {journal} {\bibinfo  {journal} {Phys. Rev. Lett.}\ }\textbf {\bibinfo
  {volume} {110}},\ \bibinfo {pages} {076402} (\bibinfo {year}
  {2013})}\BibitemShut {NoStop}%
\bibitem [{\citenamefont {Hozoi}\ \emph {et~al.}(2014)\citenamefont {Hozoi},
  \citenamefont {Gretarsson}, \citenamefont {Clancy}, \citenamefont {Jeon},
  \citenamefont {Lee}, \citenamefont {Kim}, \citenamefont {Yushankhai},
  \citenamefont {Fulde}, \citenamefont {Casa}, \citenamefont {Gog},
  \citenamefont {Kim}, \citenamefont {Said}, \citenamefont {Upton},
  \citenamefont {Kim},\ and\ \citenamefont {van~den Brink}}]{Hozoi2014}%
  \BibitemOpen
  \bibfield  {author} {\bibinfo {author} {\bibfnamefont {L.}~\bibnamefont
  {Hozoi}}, \bibinfo {author} {\bibfnamefont {H.}~\bibnamefont {Gretarsson}},
  \bibinfo {author} {\bibfnamefont {J.~P.}\ \bibnamefont {Clancy}}, \bibinfo
  {author} {\bibfnamefont {B.-G.}\ \bibnamefont {Jeon}}, \bibinfo {author}
  {\bibfnamefont {B.}~\bibnamefont {Lee}}, \bibinfo {author} {\bibfnamefont
  {K.~H.}\ \bibnamefont {Kim}}, \bibinfo {author} {\bibfnamefont
  {V.}~\bibnamefont {Yushankhai}}, \bibinfo {author} {\bibfnamefont
  {P.}~\bibnamefont {Fulde}}, \bibinfo {author} {\bibfnamefont
  {D.}~\bibnamefont {Casa}}, \bibinfo {author} {\bibfnamefont {T.}~\bibnamefont
  {Gog}}, \bibinfo {author} {\bibfnamefont {J.}~\bibnamefont {Kim}}, \bibinfo
  {author} {\bibfnamefont {A.~H.}\ \bibnamefont {Said}}, \bibinfo {author}
  {\bibfnamefont {M.~H.}\ \bibnamefont {Upton}}, \bibinfo {author}
  {\bibfnamefont {Y.-J.}\ \bibnamefont {Kim}}, \ and\ \bibinfo {author}
  {\bibfnamefont {J.}~\bibnamefont {van~den Brink}},\ }\href {\doibase
  10.1103/PhysRevB.89.115111} {\bibfield  {journal} {\bibinfo  {journal} {Phys.
  Rev. B}\ }\textbf {\bibinfo {volume} {89}},\ \bibinfo {pages} {115111}
  (\bibinfo {year} {2014})}\BibitemShut {NoStop}%
\bibitem [{\citenamefont {Liu}\ \emph {et~al.}(2012)\citenamefont {Liu},
  \citenamefont {Katukuri}, \citenamefont {Hozoi}, \citenamefont {Yin},
  \citenamefont {Dean}, \citenamefont {Upton}, \citenamefont {Kim},
  \citenamefont {Casa}, \citenamefont {Said}, \citenamefont {Gog},
  \citenamefont {Qi}, \citenamefont {Cao}, \citenamefont {Tsvelik},
  \citenamefont {van~den Brink},\ and\ \citenamefont {Hill}}]{Liu2012}%
  \BibitemOpen
  \bibfield  {author} {\bibinfo {author} {\bibfnamefont {X.}~\bibnamefont
  {Liu}}, \bibinfo {author} {\bibfnamefont {V.~M.}\ \bibnamefont {Katukuri}},
  \bibinfo {author} {\bibfnamefont {L.}~\bibnamefont {Hozoi}}, \bibinfo
  {author} {\bibfnamefont {W.-G.}\ \bibnamefont {Yin}}, \bibinfo {author}
  {\bibfnamefont {M.~P.~M.}\ \bibnamefont {Dean}}, \bibinfo {author}
  {\bibfnamefont {M.~H.}\ \bibnamefont {Upton}}, \bibinfo {author}
  {\bibfnamefont {J.}~\bibnamefont {Kim}}, \bibinfo {author} {\bibfnamefont
  {D.}~\bibnamefont {Casa}}, \bibinfo {author} {\bibfnamefont {A.}~\bibnamefont
  {Said}}, \bibinfo {author} {\bibfnamefont {T.}~\bibnamefont {Gog}}, \bibinfo
  {author} {\bibfnamefont {T.~F.}\ \bibnamefont {Qi}}, \bibinfo {author}
  {\bibfnamefont {G.}~\bibnamefont {Cao}}, \bibinfo {author} {\bibfnamefont
  {A.~M.}\ \bibnamefont {Tsvelik}}, \bibinfo {author} {\bibfnamefont
  {J.}~\bibnamefont {van~den Brink}}, \ and\ \bibinfo {author} {\bibfnamefont
  {J.~P.}\ \bibnamefont {Hill}},\ }\href {\doibase
  10.1103/PhysRevLett.109.157401} {\bibfield  {journal} {\bibinfo  {journal}
  {Phys. Rev. Lett.}\ }\textbf {\bibinfo {volume} {109}},\ \bibinfo {pages}
  {157401} (\bibinfo {year} {2012})}\BibitemShut {NoStop}%
\bibitem [{\citenamefont {Kim}\ \emph {et~al.}(2015{\natexlab{b}})\citenamefont
  {Kim}, \citenamefont {Liu}, \citenamefont {Kim}, \citenamefont {Lee},
  \citenamefont {Yao}, \citenamefont {Ho},\ and\ \citenamefont
  {Cho}}]{SWKim2015}%
  \BibitemOpen
  \bibfield  {author} {\bibinfo {author} {\bibfnamefont {S.-W.}\ \bibnamefont
  {Kim}}, \bibinfo {author} {\bibfnamefont {C.}~\bibnamefont {Liu}}, \bibinfo
  {author} {\bibfnamefont {H.-J.}\ \bibnamefont {Kim}}, \bibinfo {author}
  {\bibfnamefont {J.-H.}\ \bibnamefont {Lee}}, \bibinfo {author} {\bibfnamefont
  {Y.}~\bibnamefont {Yao}}, \bibinfo {author} {\bibfnamefont {K.-M.}\
  \bibnamefont {Ho}}, \ and\ \bibinfo {author} {\bibfnamefont {J.-H.}\
  \bibnamefont {Cho}},\ }\href {\doibase 10.1103/PhysRevLett.115.096401}
  {\bibfield  {journal} {\bibinfo  {journal} {Phys. Rev. Lett.}\ }\textbf
  {\bibinfo {volume} {115}},\ \bibinfo {pages} {096401} (\bibinfo {year}
  {2015}{\natexlab{b}})}\BibitemShut {NoStop}%
\bibitem [{\citenamefont {Ju}\ \emph {et~al.}(2013)\citenamefont {Ju},
  \citenamefont {Liu},\ and\ \citenamefont {Yang}}]{Ju2013}%
  \BibitemOpen
  \bibfield  {author} {\bibinfo {author} {\bibfnamefont {W.}~\bibnamefont
  {Ju}}, \bibinfo {author} {\bibfnamefont {G.-Q.}\ \bibnamefont {Liu}}, \ and\
  \bibinfo {author} {\bibfnamefont {Z.}~\bibnamefont {Yang}},\ }\href {\doibase
  10.1103/PhysRevB.87.075112} {\bibfield  {journal} {\bibinfo  {journal} {Phys.
  Rev. B}\ }\textbf {\bibinfo {volume} {87}},\ \bibinfo {pages} {075112}
  (\bibinfo {year} {2013})}\BibitemShut {NoStop}%
\bibitem [{\citenamefont {Streltsov}\ and\ \citenamefont
  {Khomskii}(2016)}]{Streltsov2016}%
  \BibitemOpen
  \bibfield  {author} {\bibinfo {author} {\bibfnamefont {S.~V.}\ \bibnamefont
  {Streltsov}}\ and\ \bibinfo {author} {\bibfnamefont {D.~I.}\ \bibnamefont
  {Khomskii}},\ }\href {\doibase 10.1073/pnas.1606367113} {\bibfield  {journal}
  {\bibinfo  {journal} {Proc. Natl. Acad. Sci.}\ }\textbf {\bibinfo {volume}
  {113}},\ \bibinfo {pages} {10491} (\bibinfo {year} {2016})}\BibitemShut
  {NoStop}%
\bibitem [{Note1()}]{Note1}%
  \BibitemOpen
  \bibinfo {note} {Several recent reports of pressure-induced Ir dimerizations
  in layered- and hyper-honeycomb iridates ~\cite {Veiga2017,Hermann2018}
  implies that, in edge-sharing geometries, $t_{dd}$ is almost comparable to
  $\lambda _{\protect \rm SO}$, so that relatively small pressure of $<$5 GPa
  is enough to enhance $t_{dd}$ to break the $j_{\protect \rm eff}$\protect
  \tmspace +\thinmuskip {.1667em}=\protect \tmspace +\thinmuskip {.1667em}1/2
  states in these compounds.}\BibitemShut {Stop}%
\bibitem [{\citenamefont {Blake}\ \emph {et~al.}(1998)\citenamefont {Blake},
  \citenamefont {Sloan}, \citenamefont {Vente},\ and\ \citenamefont
  {Battle}}]{Blake1998}%
  \BibitemOpen
  \bibfield  {author} {\bibinfo {author} {\bibfnamefont {G.~R.}\ \bibnamefont
  {Blake}}, \bibinfo {author} {\bibfnamefont {J.}~\bibnamefont {Sloan}},
  \bibinfo {author} {\bibfnamefont {J.~F.}\ \bibnamefont {Vente}}, \ and\
  \bibinfo {author} {\bibfnamefont {P.~D.}\ \bibnamefont {Battle}},\ }\href
  {\doibase 10.1021/cm980317m} {\bibfield  {journal} {\bibinfo  {journal}
  {Chem. Mat.}\ }\textbf {\bibinfo {volume} {10}},\ \bibinfo {pages} {3536}
  (\bibinfo {year} {1998})}\BibitemShut {NoStop}%
\bibitem [{\citenamefont {Blake}\ \emph {et~al.}(1999)\citenamefont {Blake},
  \citenamefont {Battle}, \citenamefont {Sloan}, \citenamefont {Vente},
  \citenamefont {Darriet},\ and\ \citenamefont {Weill}}]{Blake1999}%
  \BibitemOpen
  \bibfield  {author} {\bibinfo {author} {\bibfnamefont {G.~R.}\ \bibnamefont
  {Blake}}, \bibinfo {author} {\bibfnamefont {P.~D.}\ \bibnamefont {Battle}},
  \bibinfo {author} {\bibfnamefont {J.}~\bibnamefont {Sloan}}, \bibinfo
  {author} {\bibfnamefont {J.~F.}\ \bibnamefont {Vente}}, \bibinfo {author}
  {\bibfnamefont {J.}~\bibnamefont {Darriet}}, \ and\ \bibinfo {author}
  {\bibfnamefont {F.}~\bibnamefont {Weill}},\ }\href {\doibase
  10.1021/cm9807844} {\bibfield  {journal} {\bibinfo  {journal} {Chem. Mat.}\
  }\textbf {\bibinfo {volume} {11}},\ \bibinfo {pages} {1551} (\bibinfo {year}
  {1999})}\BibitemShut {NoStop}%
\bibitem [{\citenamefont {Yang}\ \emph {et~al.}(2015)\citenamefont {Yang},
  \citenamefont {Huang}, \citenamefont {Hermele}, \citenamefont {Qi},
  \citenamefont {Cao},\ and\ \citenamefont {Reznik}}]{Yang2015}%
  \BibitemOpen
  \bibfield  {author} {\bibinfo {author} {\bibfnamefont {J.-A.}\ \bibnamefont
  {Yang}}, \bibinfo {author} {\bibfnamefont {Y.-P.}\ \bibnamefont {Huang}},
  \bibinfo {author} {\bibfnamefont {M.}~\bibnamefont {Hermele}}, \bibinfo
  {author} {\bibfnamefont {T.}~\bibnamefont {Qi}}, \bibinfo {author}
  {\bibfnamefont {G.}~\bibnamefont {Cao}}, \ and\ \bibinfo {author}
  {\bibfnamefont {D.}~\bibnamefont {Reznik}},\ }\href {\doibase
  10.1103/PhysRevB.91.195140} {\bibfield  {journal} {\bibinfo  {journal} {Phys.
  Rev. B}\ }\textbf {\bibinfo {volume} {91}},\ \bibinfo {pages} {195140}
  (\bibinfo {year} {2015})}\BibitemShut {NoStop}%
\bibitem [{Note2()}]{Note2}%
  \BibitemOpen
  \bibinfo {note} {For the DFT calculations we employed the Vienna {\protect
  \it ab-initio} Simulation Package (VASP)~\cite {VASP1,VASP2}. For the details
  of the TB model and DFT calculations please refer to Supplementary
  Material.}\BibitemShut {Stop}%
\bibitem [{\citenamefont {Dudarev}\ \emph {et~al.}(1998)\citenamefont
  {Dudarev}, \citenamefont {Botton}, \citenamefont {Savrasov}, \citenamefont
  {Humphreys},\ and\ \citenamefont {Sutton}}]{DFTU}%
  \BibitemOpen
  \bibfield  {author} {\bibinfo {author} {\bibfnamefont {S.~L.}\ \bibnamefont
  {Dudarev}}, \bibinfo {author} {\bibfnamefont {G.~A.}\ \bibnamefont {Botton}},
  \bibinfo {author} {\bibfnamefont {S.~Y.}\ \bibnamefont {Savrasov}}, \bibinfo
  {author} {\bibfnamefont {C.~J.}\ \bibnamefont {Humphreys}}, \ and\ \bibinfo
  {author} {\bibfnamefont {A.~P.}\ \bibnamefont {Sutton}},\ }\href {\doibase
  10.1103/PhysRevB.57.1505} {\bibfield  {journal} {\bibinfo  {journal} {Phys.
  Rev. B}\ }\textbf {\bibinfo {volume} {57}},\ \bibinfo {pages} {1505}
  (\bibinfo {year} {1998})}\BibitemShut {NoStop}%
\bibitem [{\citenamefont {Lefran\ifmmode~\mbox{\c{c}}\else \c{c}\fi{}ois}\
  \emph {et~al.}(2016)\citenamefont {Lefran\ifmmode~\mbox{\c{c}}\else
  \c{c}\fi{}ois}, \citenamefont {Pradipto}, \citenamefont {Moretti~Sala},
  \citenamefont {Chapon}, \citenamefont {Simonet}, \citenamefont {Picozzi},
  \citenamefont {Lejay}, \citenamefont {Petit},\ and\ \citenamefont
  {Ballou}}]{Lefrancois2016}%
  \BibitemOpen
  \bibfield  {author} {\bibinfo {author} {\bibfnamefont {E.}~\bibnamefont
  {Lefran\ifmmode~\mbox{\c{c}}\else \c{c}\fi{}ois}}, \bibinfo {author}
  {\bibfnamefont {A.-M.}\ \bibnamefont {Pradipto}}, \bibinfo {author}
  {\bibfnamefont {M.}~\bibnamefont {Moretti~Sala}}, \bibinfo {author}
  {\bibfnamefont {L.~C.}\ \bibnamefont {Chapon}}, \bibinfo {author}
  {\bibfnamefont {V.}~\bibnamefont {Simonet}}, \bibinfo {author} {\bibfnamefont
  {S.}~\bibnamefont {Picozzi}}, \bibinfo {author} {\bibfnamefont
  {P.}~\bibnamefont {Lejay}}, \bibinfo {author} {\bibfnamefont
  {S.}~\bibnamefont {Petit}}, \ and\ \bibinfo {author} {\bibfnamefont
  {R.}~\bibnamefont {Ballou}},\ }\href {\doibase 10.1103/PhysRevB.93.224401}
  {\bibfield  {journal} {\bibinfo  {journal} {Phys. Rev. B}\ }\textbf {\bibinfo
  {volume} {93}},\ \bibinfo {pages} {224401} (\bibinfo {year}
  {2016})}\BibitemShut {NoStop}%
\bibitem [{\citenamefont {{Birol}}\ \emph {et~al.}(2018)\citenamefont
  {{Birol}}, \citenamefont {{Haule}},\ and\ \citenamefont
  {{Vanderbilt}}}]{Birol2018}%
  \BibitemOpen
  \bibfield  {author} {\bibinfo {author} {\bibfnamefont {T.}~\bibnamefont
  {{Birol}}}, \bibinfo {author} {\bibfnamefont {K.}~\bibnamefont {{Haule}}}, \
  and\ \bibinfo {author} {\bibfnamefont {D.}~\bibnamefont {{Vanderbilt}}},\
  }\href@noop {} {\bibfield  {journal} {\bibinfo  {journal} {ArXiv e-prints}\ }
  (\bibinfo {year} {2018})},\ \Eprint {http://arxiv.org/abs/1805.03733}
  {arXiv:1805.03733 [cond-mat.mtrl-sci]} \BibitemShut {NoStop}%
\bibitem [{\citenamefont {Mazin}\ \emph {et~al.}(2012)\citenamefont {Mazin},
  \citenamefont {Jeschke}, \citenamefont {Foyevtsova}, \citenamefont
  {Valent\'{\i}},\ and\ \citenamefont {Khomskii}}]{Mazin2012}%
  \BibitemOpen
  \bibfield  {author} {\bibinfo {author} {\bibfnamefont {I.~I.}\ \bibnamefont
  {Mazin}}, \bibinfo {author} {\bibfnamefont {H.~O.}\ \bibnamefont {Jeschke}},
  \bibinfo {author} {\bibfnamefont {K.}~\bibnamefont {Foyevtsova}}, \bibinfo
  {author} {\bibfnamefont {R.}~\bibnamefont {Valent\'{\i}}}, \ and\ \bibinfo
  {author} {\bibfnamefont {D.~I.}\ \bibnamefont {Khomskii}},\ }\href {\doibase
  10.1103/PhysRevLett.109.197201} {\bibfield  {journal} {\bibinfo  {journal}
  {Phys. Rev. Lett.}\ }\textbf {\bibinfo {volume} {109}},\ \bibinfo {pages}
  {197201} (\bibinfo {year} {2012})}\BibitemShut {NoStop}%
\bibitem [{\citenamefont {Foyevtsova}\ \emph {et~al.}(2013)\citenamefont
  {Foyevtsova}, \citenamefont {Jeschke}, \citenamefont {Mazin}, \citenamefont
  {Khomskii},\ and\ \citenamefont {Valent\'{\i}}}]{Foyevtsova2013}%
  \BibitemOpen
  \bibfield  {author} {\bibinfo {author} {\bibfnamefont {K.}~\bibnamefont
  {Foyevtsova}}, \bibinfo {author} {\bibfnamefont {H.~O.}\ \bibnamefont
  {Jeschke}}, \bibinfo {author} {\bibfnamefont {I.~I.}\ \bibnamefont {Mazin}},
  \bibinfo {author} {\bibfnamefont {D.~I.}\ \bibnamefont {Khomskii}}, \ and\
  \bibinfo {author} {\bibfnamefont {R.}~\bibnamefont {Valent\'{\i}}},\ }\href
  {\doibase 10.1103/PhysRevB.88.035107} {\bibfield  {journal} {\bibinfo
  {journal} {Phys. Rev. B}\ }\textbf {\bibinfo {volume} {88}},\ \bibinfo
  {pages} {035107} (\bibinfo {year} {2013})}\BibitemShut {NoStop}%
\bibitem [{\citenamefont {Veiga}\ \emph {et~al.}(2017)\citenamefont {Veiga},
  \citenamefont {Etter}, \citenamefont {Glazyrin}, \citenamefont {Sun},
  \citenamefont {Escanhoela}, \citenamefont {Fabbris}, \citenamefont
  {Mardegan}, \citenamefont {Malavi}, \citenamefont {Deng}, \citenamefont
  {Stavropoulos}, \citenamefont {Kee}, \citenamefont {Yang}, \citenamefont {van
  Veenendaal}, \citenamefont {Schilling}, \citenamefont {Takayama},
  \citenamefont {Takagi},\ and\ \citenamefont {Haskel}}]{Veiga2017}%
  \BibitemOpen
  \bibfield  {author} {\bibinfo {author} {\bibfnamefont {L.~S.~I.}\
  \bibnamefont {Veiga}}, \bibinfo {author} {\bibfnamefont {M.}~\bibnamefont
  {Etter}}, \bibinfo {author} {\bibfnamefont {K.}~\bibnamefont {Glazyrin}},
  \bibinfo {author} {\bibfnamefont {F.}~\bibnamefont {Sun}}, \bibinfo {author}
  {\bibfnamefont {C.~A.}\ \bibnamefont {Escanhoela}}, \bibinfo {author}
  {\bibfnamefont {G.}~\bibnamefont {Fabbris}}, \bibinfo {author} {\bibfnamefont
  {J.~R.~L.}\ \bibnamefont {Mardegan}}, \bibinfo {author} {\bibfnamefont
  {P.~S.}\ \bibnamefont {Malavi}}, \bibinfo {author} {\bibfnamefont
  {Y.}~\bibnamefont {Deng}}, \bibinfo {author} {\bibfnamefont {P.~P.}\
  \bibnamefont {Stavropoulos}}, \bibinfo {author} {\bibfnamefont {H.-Y.}\
  \bibnamefont {Kee}}, \bibinfo {author} {\bibfnamefont {W.~G.}\ \bibnamefont
  {Yang}}, \bibinfo {author} {\bibfnamefont {M.}~\bibnamefont {van
  Veenendaal}}, \bibinfo {author} {\bibfnamefont {J.~S.}\ \bibnamefont
  {Schilling}}, \bibinfo {author} {\bibfnamefont {T.}~\bibnamefont {Takayama}},
  \bibinfo {author} {\bibfnamefont {H.}~\bibnamefont {Takagi}}, \ and\ \bibinfo
  {author} {\bibfnamefont {D.}~\bibnamefont {Haskel}},\ }\href {\doibase
  10.1103/PhysRevB.96.140402} {\bibfield  {journal} {\bibinfo  {journal} {Phys.
  Rev. B}\ }\textbf {\bibinfo {volume} {96}},\ \bibinfo {pages} {140402}
  (\bibinfo {year} {2017})}\BibitemShut {NoStop}%
\bibitem [{\citenamefont {Hermann}\ \emph {et~al.}(2018)\citenamefont
  {Hermann}, \citenamefont {Altmeyer}, \citenamefont {Ebad-Allah},
  \citenamefont {Freund}, \citenamefont {Jesche}, \citenamefont {Tsirlin},
  \citenamefont {Hanfland}, \citenamefont {Gegenwart}, \citenamefont {Mazin},
  \citenamefont {Khomskii}, \citenamefont {Valent\'{\i}},\ and\ \citenamefont
  {Kuntscher}}]{Hermann2018}%
  \BibitemOpen
  \bibfield  {author} {\bibinfo {author} {\bibfnamefont {V.}~\bibnamefont
  {Hermann}}, \bibinfo {author} {\bibfnamefont {M.}~\bibnamefont {Altmeyer}},
  \bibinfo {author} {\bibfnamefont {J.}~\bibnamefont {Ebad-Allah}}, \bibinfo
  {author} {\bibfnamefont {F.}~\bibnamefont {Freund}}, \bibinfo {author}
  {\bibfnamefont {A.}~\bibnamefont {Jesche}}, \bibinfo {author} {\bibfnamefont
  {A.~A.}\ \bibnamefont {Tsirlin}}, \bibinfo {author} {\bibfnamefont
  {M.}~\bibnamefont {Hanfland}}, \bibinfo {author} {\bibfnamefont
  {P.}~\bibnamefont {Gegenwart}}, \bibinfo {author} {\bibfnamefont {I.~I.}\
  \bibnamefont {Mazin}}, \bibinfo {author} {\bibfnamefont {D.~I.}\ \bibnamefont
  {Khomskii}}, \bibinfo {author} {\bibfnamefont {R.}~\bibnamefont
  {Valent\'{\i}}}, \ and\ \bibinfo {author} {\bibfnamefont {C.~A.}\
  \bibnamefont {Kuntscher}},\ }\href {\doibase 10.1103/PhysRevB.97.020104}
  {\bibfield  {journal} {\bibinfo  {journal} {Phys. Rev. B}\ }\textbf {\bibinfo
  {volume} {97}},\ \bibinfo {pages} {020104} (\bibinfo {year}
  {2018})}\BibitemShut {NoStop}%
\bibitem [{\citenamefont {Kresse}\ and\ \citenamefont
  {Furthm\"uller}(1996)}]{VASP1}%
  \BibitemOpen
  \bibfield  {author} {\bibinfo {author} {\bibfnamefont {G.}~\bibnamefont
  {Kresse}}\ and\ \bibinfo {author} {\bibfnamefont {J.}~\bibnamefont
  {Furthm\"uller}},\ }\href {\doibase 10.1103/PhysRevB.54.11169} {\bibfield
  {journal} {\bibinfo  {journal} {Phys. Rev. B}\ }\textbf {\bibinfo {volume}
  {54}},\ \bibinfo {pages} {11169} (\bibinfo {year} {1996})}\BibitemShut
  {NoStop}%
\bibitem [{\citenamefont {Kresse}\ and\ \citenamefont {Joubert}(1999)}]{VASP2}%
  \BibitemOpen
  \bibfield  {author} {\bibinfo {author} {\bibfnamefont {G.}~\bibnamefont
  {Kresse}}\ and\ \bibinfo {author} {\bibfnamefont {D.}~\bibnamefont
  {Joubert}},\ }\href {\doibase 10.1103/PhysRevB.59.1758} {\bibfield  {journal}
  {\bibinfo  {journal} {Phys. Rev. B}\ }\textbf {\bibinfo {volume} {59}},\
  \bibinfo {pages} {1758} (\bibinfo {year} {1999})}\BibitemShut {NoStop}%
\bibitem [{\citenamefont {Perdew}\ \emph {et~al.}(2008)\citenamefont {Perdew},
  \citenamefont {Ruzsinszky}, \citenamefont {Csonka}, \citenamefont {Vydrov},
  \citenamefont {Scuseria}, \citenamefont {Constantin}, \citenamefont {Zhou},\
  and\ \citenamefont {Burke}}]{PBE}%
  \BibitemOpen
  \bibfield  {author} {\bibinfo {author} {\bibfnamefont {J.~P.}\ \bibnamefont
  {Perdew}}, \bibinfo {author} {\bibfnamefont {A.}~\bibnamefont {Ruzsinszky}},
  \bibinfo {author} {\bibfnamefont {G.~I.}\ \bibnamefont {Csonka}}, \bibinfo
  {author} {\bibfnamefont {O.~A.}\ \bibnamefont {Vydrov}}, \bibinfo {author}
  {\bibfnamefont {G.~E.}\ \bibnamefont {Scuseria}}, \bibinfo {author}
  {\bibfnamefont {L.~A.}\ \bibnamefont {Constantin}}, \bibinfo {author}
  {\bibfnamefont {X.}~\bibnamefont {Zhou}}, \ and\ \bibinfo {author}
  {\bibfnamefont {K.}~\bibnamefont {Burke}},\ }\href {\doibase
  10.1103/PhysRevLett.100.136406} {\bibfield  {journal} {\bibinfo  {journal}
  {Phys. Rev. Lett.}\ }\textbf {\bibinfo {volume} {100}},\ \bibinfo {pages}
  {136406} (\bibinfo {year} {2008})}\BibitemShut {NoStop}%
\end{thebibliography}%

\clearpage

\renewcommand{\thefigure}{S\arabic{figure}}

\subsection{Computational details}

\subsubsection{Ab-initio simulations}

For our {\it ab-initio} simulations, we first assume a crystal structure with no disorder in Cu positions, that Cu are located at one of the three equivalent prismatic facial positions as shown in Fig.~2 in the main text. Further, no Cu-Ir intermixing is considered. The unit cell contains 6 formula units, i.e. three Ir-Cu chains. It should be commented that, for a better reproduction of the experimentally measured spectra one may need to take an average over {\it ab-initio} calculations from all possible orderings of Cu to restore the three-fold rotation symmetry along the chain direction. For a qualitative understanding of the high-energy spectra, however, the current result from the choice of Cu ordering shown in Fig.~2 in the main text seems to be enough.

The Vienna {\it ab-initio} Simulation Package (VASP), which uses the projector-augmented wave (PAW) basis set~\cite{VASP1,VASP2}, is employed for structural optimizations and electronic structure calculations. 400 eV and a $\Gamma$-centered 3x3x3 $k$-point grid are used for the plane wave energy cutoff and the $k$-point sampling, respectively. A revised Perdew-Burke-Ernzerhof generalized gradient approximation for crystalline solids (PBEsol) is chosen for the exchange-correlation functional~\cite{PBE}, and $10^{-2}$ eV/\AA~of force criterion was used for the structural optimization. A simplified rotationally-invariant form of DFT+$U$ method~\cite{DFTU} is employed to treat the on-site Coulomb interaction both at Cu and Ir sites with the spin-orbit coupling (SOC) included.

\subsubsection{Tight-binding model}

Assuming the three-fold rotational symmetry along the chain direction ({\it i.e.} ignoring the IrO$_6$ octahedral distortions other than the trigonal one), and also assuming the Ir-Ir hybridizations are dominating those from Ir-Cu (this choice is justified by DFT calculations which show small hybridization between Ir and Cu $d$-orbitals), the three-site tight-binding model can be written as follows,
\begin{align*}
H_{\rm TB} = \left(
\begin{array}{ccc|ccc|ccc}
0 & \Delta_{\rm tri} & \Delta_{\rm tri} &
t_1 & t_2 & t_2 &
0 & 0 & 0 \\
\Delta_{\rm tri} & 0 & \Delta_{\rm tri} &
t_2 & t_1 & t_2 &
0 & 0 & 0 \\
\Delta_{\rm tri} & \Delta_{\rm tri} & 0 &
t_2 & t_2 & t_1 &
0 & 0 & 0 \\ \hline
t_1 & t_2 & t_2 &
\epsilon_{\rm on} & \Delta_{\rm tri} & \Delta_{\rm tri} &
t_1 & t_2 & t_2 \\
t_2 & t_1 & t_2 &
\Delta_{\rm tri} & \epsilon_{\rm on} & \Delta_{\rm tri} &
t_2 & t_1 & t_2 \\
t_2 & t_2 & t_1 &
\Delta_{\rm tri} & \Delta_{\rm tri} & \epsilon_{\rm on} & 
t_2 & t_2 & t_1 \\ \hline
0 & 0 & 0 &
t_1 & t_2 & t_2 &
0 & \Delta_{\rm tri} & \Delta_{\rm tri} \\
0 & 0 & 0 &
t_2 & t_1 & t_2 &
\Delta_{\rm tri} & 0 & \Delta_{\rm tri} \\
0 & 0 & 0 &
t_2 & t_2 & t_1 &
\Delta_{\rm tri} &  \Delta_{\rm tri} & 0
\end{array} \right),
\end{align*}
where $t_1$ and $t_2$ are nearest-neighbor hopping integrals between the same and different kinds of Ir $t_{\rm 2g}$ orbitals respectively, $\Delta_{\rm tri}$ is the on-site trigonal crystal field within the $t_{\rm 2g}$ complexes, and $\epsilon_{\rm on}$ is the on-site energy difference between the central and lateral Ir. The projected density of states in Fig.~3(b) in the main text are calculated by choosing $t_1$ = -0.43 eV, $t_2$ = -0.20 eV, $\Delta_{\rm tri}$ = -0.12 eV, and $\epsilon_{\rm on}$ = -0.3 eV, which yield qualitative agreement with the {\it ab-initio} results as discussed in the main text. Note that, if we take the unitary transform to the trigonal basis set ($a_{\rm 1g}$ and $e'_{\rm g}$ states), the $\sigma$-like overlap between the $a_{\rm 1g}$ states becomes $t_1+2t_2$ = -0.83 eV, twice larger than the SOC of Ir ($\lambda_{\rm SO}$ $\simeq$ 0.4 eV), supporting the molecular-orbital scenario.

\subsubsection{$U_{\rm Ir}$ dependence of the electronic structure}

\begin{figure*}
\includegraphics[width=0.90\textwidth]{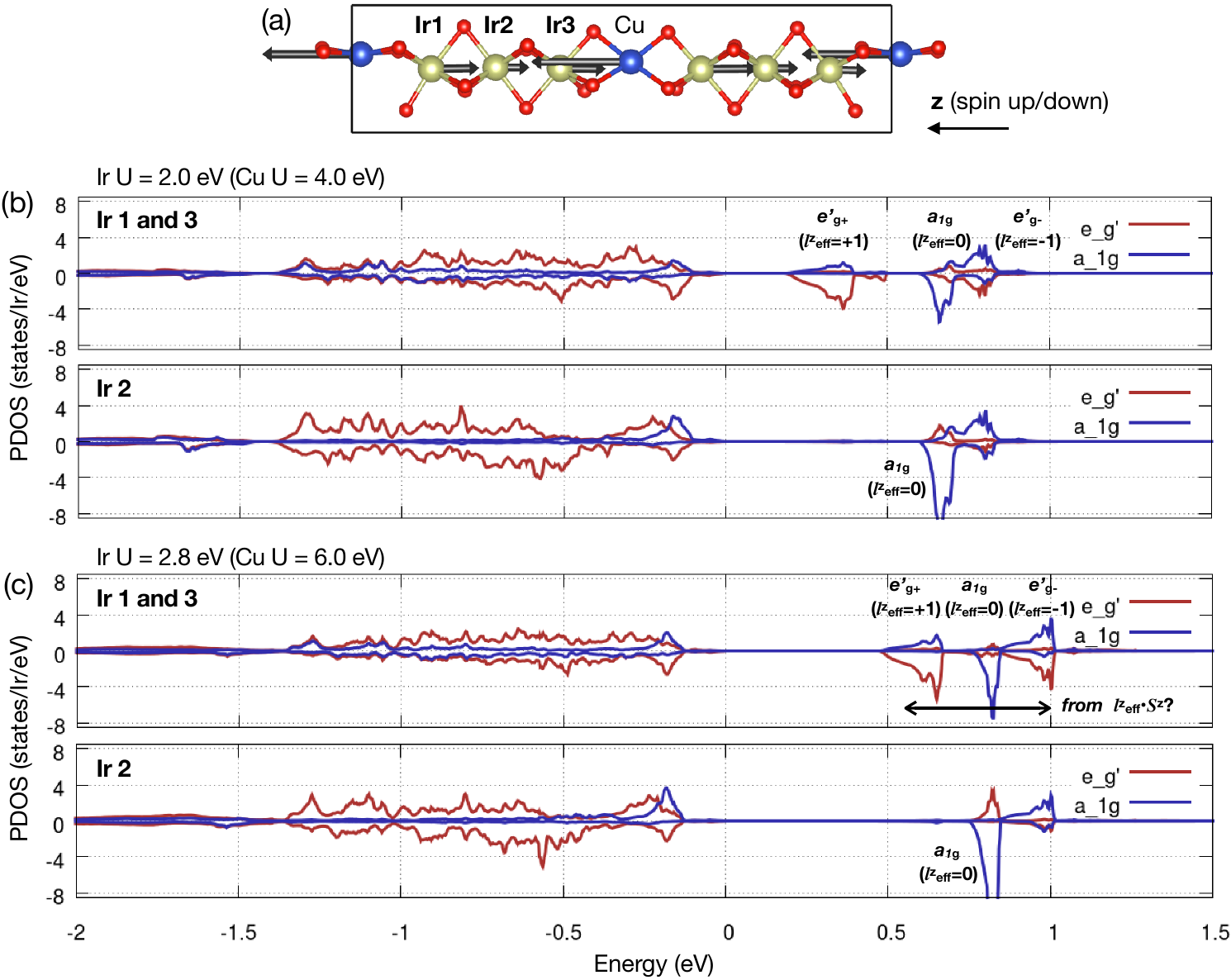}
\caption{\label{fig:FigS2} (a) The ground state magnetic configuration in a Cu-Ir chain, where the black arrows depict the size and direction of the spin moments at Cu and Ir sites. (b,c) $a_{\rm 1g}$/$e'_{\rm g}$-projected DOS with (b) ($U_{\rm Ir}, U_{\rm Cu}$) = (2, 4)eV and (c) (2.8, 6)eV.}
\end{figure*}

In the computations of projected DOS presented in the main text, a collinear magnetic configuration illustrated in Fig.~\ref{fig:FigS2}(a) is employed. Note that in our calculations with various trial magnetic configurations, the Ir spin moments on a Ir trimer show a collective motion and favor the direction parallel to the Cu-Ir chain, consistent with the effect of MO formation and the role SOC discussed in the main text. Also note that the one depicted in Fig.~\ref{fig:FigS2}(a) is the ground state configuration for our choice of Cu-Ir chain structure [see Fig.~1 in the main text], while the energetics and the ground state configuration may depend on a different choice of Cu order. 

In our DFT+$U$ calculations we choose $U$ value for Cu (denoted as $U_{\rm Cu}$) to be 4$\sim$6 eV. With this choice of $U_{\rm Cu}$ the Cu $e_{\rm g}$ orbital character is almost absent near the Fermi level, so that the electronic structure near the Fermi level is mostly determined by the Ir $t_{\rm 2g}$ states. As discussed in the main text, the combination of Ir SOC and $U_{\rm Ir}$ induces the three-peak structure as observed in the Raman measurement. In the main text we employ $U_{\rm Ir}$ = 2.8 eV, and the three peak structure does not qualitatively change in the range of $2 < U_{\rm Ir} < 3$ eV. Fig.~\ref{fig:FigS2}(b) and (c) show the projected DOS with  ($U_{\rm Ir}, U_{\rm Cu}$) = (2, 4)eV and (2.8, 6)eV respectively. It can be seen that smaller value of $U_{\rm Ir}$ induces smaller charge gap, but the three-peak structure remains almost unchanged. the spacing between the three peaks in the upper Hubbard bands can be affected in a quantitative way with a different choice of  $U_{\rm Ir}$.

\subsection{Experimental details}

\subsubsection{Material preparation}

Single crystals of Ba$_5$CuIr$_3$O$_{12}$ are grown by flux method; details of growth and characterization will be published separately. This material has a trigonal structure (space group $P3c1$, No.158; point group $C_{3v}$)~\cite{Blake1998,Blake1999}. The sample is polished with a lapping film (1 micrometer, Buehler) and is subsequently annealed at 650~$^\circ$C in air to remove residual strain. Its annealed (100) crystallographic surface is then used for Raman measurements.

\subsubsection{Raman scattering}

Raman-scattering measurements are performed in a quasi-back-scattering optical setup. The 476.2\,nm line from a Kr$^+$ ion laser is for excitation. Incident light with $\sim$10\,mW power is focused to a 50$\times$100\,$\mu$m$^{2}$ spot on the (100) crystallographic surface. Environmental temperature of 20\,K is achieved in a helium-gas-cooled cryostat, and the laser heating is assumed to be 0.5\,K$\slash$\,mW. We use a custom triple-grating spectrometer and a liquid-nitrogen-cooled charge-coupled device (CCD) detector for collection and analysis of the scattered light. The data are corrected for the system background and the spectral response. The measured scattering intensity $I(\omega,T)$ is related to the Raman response $\chi''(\omega,T)$ by $I(\omega,T)=[1+n(\omega,T)]\chi''(\omega,T)$, where $n$ is the Bose factor, $\omega$ is Raman shift and $T$ is temperature.

\subsection{Low-energy Raman spectrum}

\begin{figure}
\includegraphics[width=0.40\textwidth]{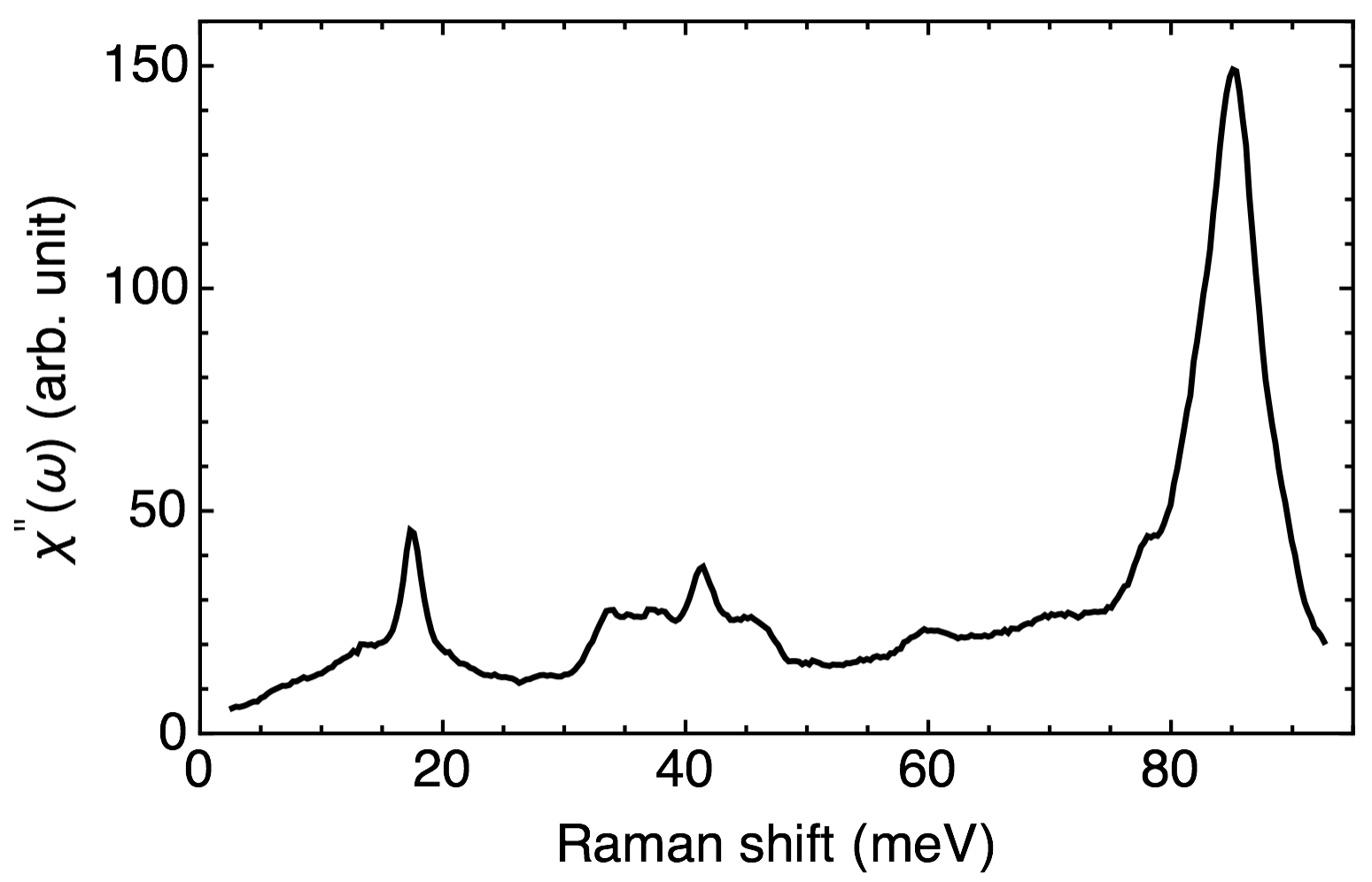}
\caption{\label{fig:FigS1}Raman response $\chi''(\omega)$ of Ba$_5$CuIr$_3$O$_{12}$ at 25\,K.}
\end{figure}

Fig.~\ref{fig:FigS1} shows the low-energy Raman spectrum of Ba$_5$CuIr$_3$O$_{12}$ at 25\,K. The sharp features at 17, 41 and 84\,meV are identified as phonon modes. The 17\,meV mode primarily involves motion of Ir atoms, and the 84\,meV mode is derived from vibration of O atoms. The broad feature centered at 40\,meV could be a bundle of phonon modes related to the motion of Ba atoms. The two weak features at 60 and 78\,meV are also likely phonon modes. 

\end{document}